\documentclass[11pt, a4paper, onecolumn]{article}

\usepackage{graphicx} 
\usepackage{graphicx,subfigure}
\usepackage{amssymb} 
\usepackage{amsmath}
\usepackage{algorithm}
\usepackage{algpseudocode}
\usepackage{tikz}
\usepackage{booktabs}

\usepackage{tcolorbox}
\newenvironment{mybox}[1]{%
\tcolorbox[float=t, savedelimiter=mybox,
savelowerto=\jobname_bspsave2.tex,
lowerbox=ignored,boxrule=1pt,
colback=white,colframe=blue!45,fonttitle=\bfseries,title=#1]}%
{\endtcolorbox}
\usepackage{makeidx}        
\usepackage{natbib} 
\makeindex                  
\usepackage{float}

\usepackage[left=1.6cm, right=1.6cm, top=1.6cm, bottom=2cm]{geometry}

\newtheorem{definition}{Definition}
\newtheorem{assumption}{Assumption}

\newtheorem{theorem}{Theorem}
\newtheorem{corollary}{Corollary}

\newtheorem{lemma}{Lemma}

\def\ba{\begin{aligned}}
\def\ea{\end{aligned}}

\newcommand{\hbeta}{\beta}

\newcommand{\ub}{\mathbf{u}}
\newcommand{\Vb}{\mathbf{V}}
\newcommand{\xb}{\mathbf{x}}
\newcommand{\wb}{\mathbf{w}}

\newcommand{\Mb}{\mathbf{M}}

\newcommand{\Kb}{\mathbf{K}}
\newcommand{\qse}{q_{\rm se}}
\newcommand{\qac}{q_{\rm ac}}
\newcommand{\Ab}{\mathbf{A}}
\newcommand{\Bb}{\mathbf{B}}

\title{\bf Quantum Encrypted Control of Networked Systems}

\author{%
    Zihao Ren,
    Daniel Quevedo,
    Salah Sukkarieh, and
    Guodong Shi\thanks{Z. Ren, S. Sukkarieh, and G. Shi are with the Australian Centre for Robotics, School of Aerospace, Mechanical and Mechatronic Engineering, The University of Sydney, NSW 2006, Australia (Email: zren0735@uni.sydney.edu.au, salah.sukkarieh@sydney.edu.au, guodong.shi@sydney.edu.au). D. Quevedo is with the School of Electrical and Computer Engineering, The University of Sydney, NSW 2006, Australia (Email: daniel.quevedo@sydney.edu.au). A preliminary work has been submitted to the 23rd IFAC World Congress, August 23-28, 2026, BEXCO, Busan, Republic of Korea. This work was supported by the Australian Research Council under Grant DP190103615, Grant LP210200473, and Grant DP230101014, and by the Faculty of Engineering's Breakthrough Project at The University of Sydney.}
}

\date{}
\begin{document}

\maketitle

\begin{abstract}
Encrypted control has been extensively studied to ensure the confidentiality of system states and control inputs for networked control systems. This paper presents a computationally efficient encrypted control framework for networked systems enabled by quantum communication. A quantum channel between sensors and actuators is used to generate identical secret keys, whose security is further enhanced through quantum key distribution. These keys enable lightweight encryption and decryption while preserving confidentiality and control accuracy. We develop a novel encryption--decryption architecture for state-feedback control of linear systems based on quantum keys, and characterize the impact of quantum state errors on closed-loop stability. In particular, we establish the existence of a critical threshold on intrinsic quantum noise below which stability is guaranteed. In contrast to classical encrypted control schemes, which may collapse under a single key-bit error, the proposed quantum encrypted control exhibits strong robustness to key imperfections. We further adopt quantization techniques to address the scenarios with limited communication bits in practical situations, and implement privacy protection for quantum keys based on a stochastic quantizer. These results demonstrate that integrating quantum technologies into control systems in a nontrivial and principled manner, even at their current level of maturity, can yield substantial performance gains in reducing computational complexity and improving resilience to key errors while ensuring security against multiple eavesdropping sources.
\end{abstract}


\vspace{1cm}

\section{Introduction}

Networked control systems are systems where the sensor, controller, and  actuator are connected via communication networks. Driven by the goal of achieving greater scalability and enhanced performance, networked  systems have found extensive use in applications that leverage distributed and cloud-based computing, e.g., \cite{NCS1,Daniel_CSM}. Cyber-physical security risks in such systems are becoming increasingly critical, as information leakage or manipulation in the cyber layer can lead to catastrophic failures in the physical domain \cite{attack_mode, ECON2, EAV1, EAV0}.  To mitigate information leakage risks in networked systems, encrypted control has received extensive attention, with the goal of protecting against eavesdropping both along communication channels and within the controllers themselves \cite{Con_HE1,Con_HE2,Con_HE3,A_AES,PAR_CON}.

In \cite{Con_HE1}, asymmetric encryption schemes such as  RSA \cite{RSA} and Elgamal \cite{ELE}, were applied to encrypted control systems, where only the actuator held the decryption key, ensuring that eavesdroppers could not decrypt the ciphertext.  
 Notably, by leveraging the  multiplicatively homomorphic
 properties of asymmetric encryption, which allowed multiple computations to be performed directly on ciphertext, this system enabled the controller to process encrypted data without requiring access to the decryption key, thereby providing resistance to eavesdropping in the controller. Furthermore, 
based on the additively homomorphic encryption scheme, Parliar \cite{PAR}, which supported both additive and scalar multiplication operations on ciphertext, \cite{PAR_CON} introduced a simpler structure for linear encrypted control. Furthermore, studies in \cite{Con_HE2, Con_HE3} investigated the use of encryption methods with stronger homomorphic properties, focusing on the
 trade-offs between computational efficiency and control
 performance in system design. In addition, \cite{A_AES} used symmetric encryption \cite{KDE1,KDE2,AES1} with a shared key between the sensor, actuator, and controller, effectively reducing the computational complexity of the system. This approach also mitigated the risk of eavesdropping on a single controller by distributing information across two distinct controllers.
Beyond fundamental control design, research on encrypted control expanded to tackle issues such as communication quantization in encrypted control \cite{quantization,quantization2}, as well as its application in model predictive control (MPC) \cite{MPC} and optimization-based scenarios \cite{optimi,optimi2}. A comprehensive survey of these advancements can be found in \cite{Daniel_CSM}.

Along with the strong security guarantees brought by asymmetrically encrypted control, its practical implementation faces two major challenges.
 First, since it often involves operations such as modular exponentiation with large integers and elliptic-curve point multiplication, it tends to have high computational complexity, which is particularly critical in networked control systems, where excessive processing overhead can induce undesirable time delays. Second, asymmetric encryption is highly sensitive to key errors. This is because the unbiased decryption of asymmetric encryption mainly relies on certain mathematical formulas involving modulus and exponentiation, e.g., encryption schemes RSA and ElGamal are based on Euler’s theorem. If deviations occur during key transmission, decrypting with an incorrect key will result in entirely incorrect information even if only a single bit of either the public or private key is corrupted.

In this paper, we propose a quantum encrypted control scheme (\(\mathsf{QEC}\)) in which quantum keys are distributed through a quantum channel between the sensor and the actuator.  In  \(\mathsf{QEC}\), the controller directly computes on the ciphertext of the state to derive the ciphertext of the control input without decrypting the state, while the sensor and actuator encrypt the state and decrypt the control input using quantum keys, respectively. Compared to symmetrically encrypted control,  \(\mathsf{QEC}\)  eliminates the need to distribute keys to the controller, thereby reducing the risk of eavesdropping in the controller. Asymmetrically encrypted control relies on computationally hard problems for security and thus incurs high computational complexity. In contrast, the security of \(\mathsf{QEC}\) is grounded in quantum key distribution, which lends itself to simple implementations which involve only low-complexity operations. Consequently, the computational overhead is substantially reduced.
For the proposed \(\mathsf{QEC}\) realization, we analyze the convergence conditions under quantum key errors. We also examine the scenario in which communication network bandwidth is limited. Through stochastic quantization, we achieve both privacy protection of quantum keys and efficient adaptation to bandwidth constraints.
The main contributions of this work are summarized as follows:
\begin{itemize}
    \item[(i)] We develop a lightweight and computationally efficient exponential–logarithm realization of \(\mathsf{QEC}\) that prevents eavesdropping from both the communication channel and the controller. 
    \item[(ii)] We establish a theoretical framework to characterize robustness against quantum-bit-induced key deviations. By analyzing the convergence of the encrypted dynamics under perturbed keys, we demonstrate the resilience of \(\mathsf{QEC}\) to key mismatch.
    \item[(iii)] We incorporate stochastic quantization techniques to address the practical scenario of limited communication bandwidth. This approach also preserves the privacy of the quantum keys, indicating a quantum-classical confidentiality feedback loop. Furthermore, we establish a trade-off between quantization error and privacy protection strength, ensuring both security and control performance of the system.
\end{itemize}

The remainder of this paper is organized as follows.  
Section~\ref{sec.pro} reviews networked control systems and classical symmetric and asymmetric encrypted control schemes, and refines the problem of interest. 
Section~\ref{sec.qec} presents the proposed \(\mathsf{QEC}\) and its specific algorithmic realization. Section~\ref{sec.mat} establishes the convergence conditions under key mismatch.  Section~\ref{sec.quan} applies a stochastic quantizer to  \(\mathsf{QEC}\) for a fully digital implementation. Section \ref{DP} establishes the trade-off between quantization error and the privacy protection strength of the quantum key from the classical quantization. 
Section~\ref{sec.num} reports numerical results and comparisons with existing encryption-based control methods.  
Finally, Section~\ref{sec.con} concludes the paper. All proofs are collected in the Appendices. 

Preliminary results were submitted for presentation at the 23rd IFAC World Congress. Compared to the conference version, this paper further incorporates quantization techniques, enabling the proposed algorithm to be applied in scenarios with limited communication bits. A stochastic quantizer is introduced to achieve privacy protection for the quantum keys, and the trade-off between privacy protection and quantization error is derived. Additionally, the journal version has been significantly improved, with enhanced clarity and more detailed analysis.

\noindent
{\bf Notations:}  
The symbol \( \otimes \) denotes the Kronecker product between quantum states or operators.
The notation $\overline{b_{w-1}b_{w-2}...b_1b_0}\in\{0,1\}_w$ denotes a string of $w$-bit binary numbers for $b_{w-1},...,b_0\in\{0,1\}$.  
For a random variable $x$, its expectation and variance are denoted by $\mathsf{E}(x)$ and $\mathsf{D}(x)$, respectively.

\section{Preliminaries and Problem Definition}
\label{sec.pro}
We review fundamental concepts in networked control systems and encrypted control, and formulate the problem of interest by highlighting a key gap in the confidentiality–robustness–efficiency trilemma.

\subsection{Networked Systems}

\begin{figure}[t]
    \centering
\includegraphics[width=9cm]{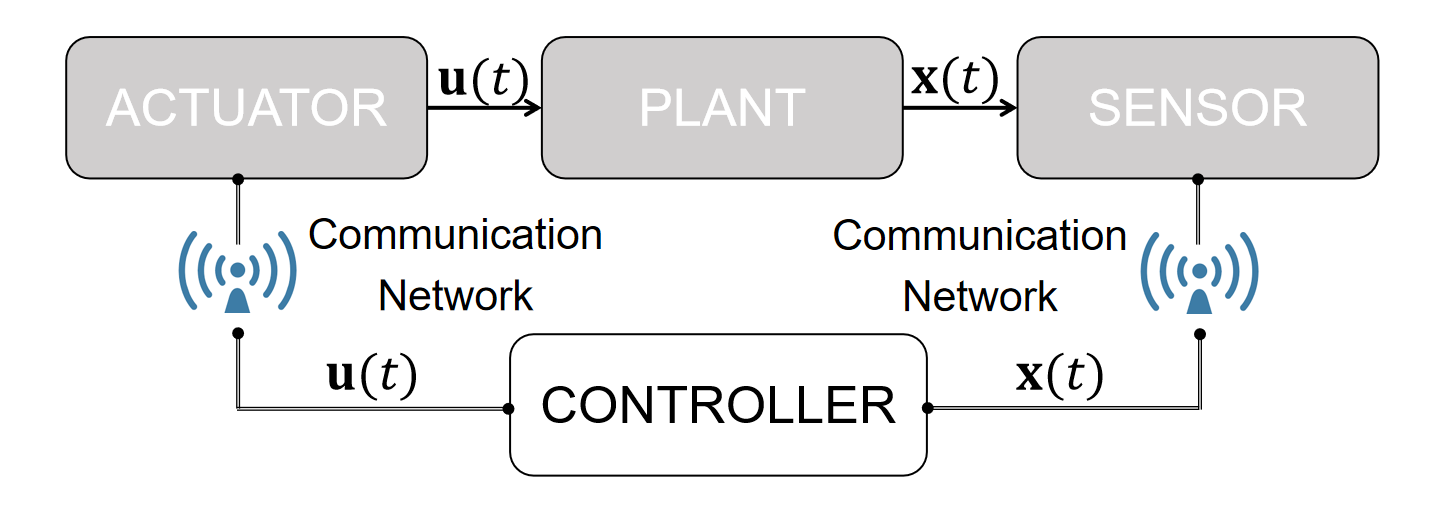}  

\caption{A networked control system.}
    \label{fig.illu_net}
\end{figure}

We consider a networked system that operates over communication channels between the sensor and controller, and between the controller and actuator, as illustrated in Fig. \ref{fig.illu_net}.  In the most basic setting,  the plant dynamics are governed by a linear system described by 
\begin{equation}
\label{eq:control}
\xb(t+1) = \Ab \xb(t) + \Bb \ub(t),
\end{equation}
for $t \in \mathbb{N}$, where $\xb(t) \in \mathbb{R}^n$ denotes the plant state, $\ub(t) \in \mathbb{R}^m$ is the control input, $\Ab \in \mathbb{R}^{n\times n}$ is the system matrix, and $\Bb \in \mathbb{R}^{n\times m}$ is the input matrix; and  a simple state-feedback control law 
\begin{equation}
\label{eq:controllaw}
\ub(t) = \Kb \xb(t),
\end{equation}
is adopted, where $\Kb \in \mathbb{R}^{m\times n}$ is the   gain matrix. 
Without loss of generality, we let $m=1$ for simplicity.

\medskip
 
\subsection{Encryption Algorithms}
 
Encryption is a technique used to prevent eavesdropping in communication between the sender and the receiver. When the sender is ready to transmit information, i.e., the plaintext \(x \in \mathbb{R}\), it uses specific parameters, referred to as the encryption key \(k_{\rm Enc} \in \mathbb{K}\), to encrypt \(x\) using an encryption function
$
f_{\rm Enc}: \mathbb{R} \times \mathbb{K} \rightarrow \mathbb{C}.
$
Here \(\mathbb{K}\) is the key space, and \(\mathbb{C}\) denotes the ciphertext space. The result of this encryption is the ciphertext
$
\tilde{x} = f_{\rm Enc}(x, k_{\rm Enc})
$
 which is then transmitted to the receiver, who possesses the decryption key \(k_{\rm Dec}\). Using a corresponding decryption function
$
f_{\rm Dec}: \mathbb{C} \times \mathbb{K} \rightarrow \mathbb{R},
$
the receiver can recover the original plaintext
$
x = f_{\rm Dec}(\tilde{x}, k_{\rm Dec}).
$
 A key property of this scheme is that, without access to  $k_{\rm Dec}$, an eavesdropper accessing $\tilde{x}$ cannot recover the original plaintext \(x\), thus ensuring the secure transmission of \(x\).




Symmetric encryption is an encryption scheme in which the same secret key $k_{\rm s}\in\mathbb{K}$ is used for both encryption and decryption \cite{KDE1,KDE2,AES1}, i.e., $k_{\rm Enc}=k_{\rm Dec}=k_{\rm s}$. 
In practice, a  key $k_{\rm s}$ is securely shared between the sender and the receiver, by which the sender encrypts $x$ in
$
\tilde{x} = f_{\rm Enc}(x, k_{\rm s}),
$
and the receiver decrypts $\tilde{x}$ in
$
x = f_{\rm Dec}(\tilde{x}, k_{\rm s}).
$ The security of symmetric encryption depends primarily on secure key distribution. For existing symmetric encryption methods, as long as the key is not compromised, an eavesdropper cannot recover the plaintext from the ciphertext.



In  asymmetric encryption, the encryption and decryption operations use different keys \cite{RSA,ELE,PAR,ECC,FHE3}. The encryption key \(k_{\rm Enc} \in \mathbb{K}\) is made publicly available, referred to as the public key $k_{\rm p}=k_{\rm Enc}$, and is used by the sender to encrypt \(x\) as
$
\tilde{x} = f_{\rm Enc}(x, k_{\rm p}).
$
On the other hand, the decryption key \(k_{\rm Dec} \in \mathbb{K}\) is kept secret by the receiver, known as the private key $k_{\rm s}=k_{\rm Dec}$, which allows the receiver to decrypt \(\tilde{x}\) as
$
x = f_{\rm Dec}(\tilde{x}, k_{\rm s}).
$ The security of asymmetric encryption relies on the computational difficulty of certain mathematical problems. Due to computational difficulty, even if an eavesdropper knows the public encryption key $k_{\rm p}$, the encryption function $f_{\rm Enc}$, and the ciphertext $\tilde{x} = f_{\rm Enc}(x, k_{\rm p})$, it cannot reversely compute the plaintext $x$.
 For example, schemes such as RSA \cite{RSA}, ElGamal \cite{ELE}, Paillier \cite{PAR} and ECC \cite{ECC} are based on the hardness of integer factorization and discrete logarithm problems. However, with the advent of quantum computing, Shor’s algorithm \cite{SHOR} can efficiently solve these problems, enabling an eavesdropper with a quantum computer to obtain the data encrypted by these schemes.

\subsection{Encrypted Control}

\begin{figure*}[t]
    \centering
       \begin{subfigure}[Symmetrically Encrypted Control]{	
	\includegraphics[width=8.5cm]{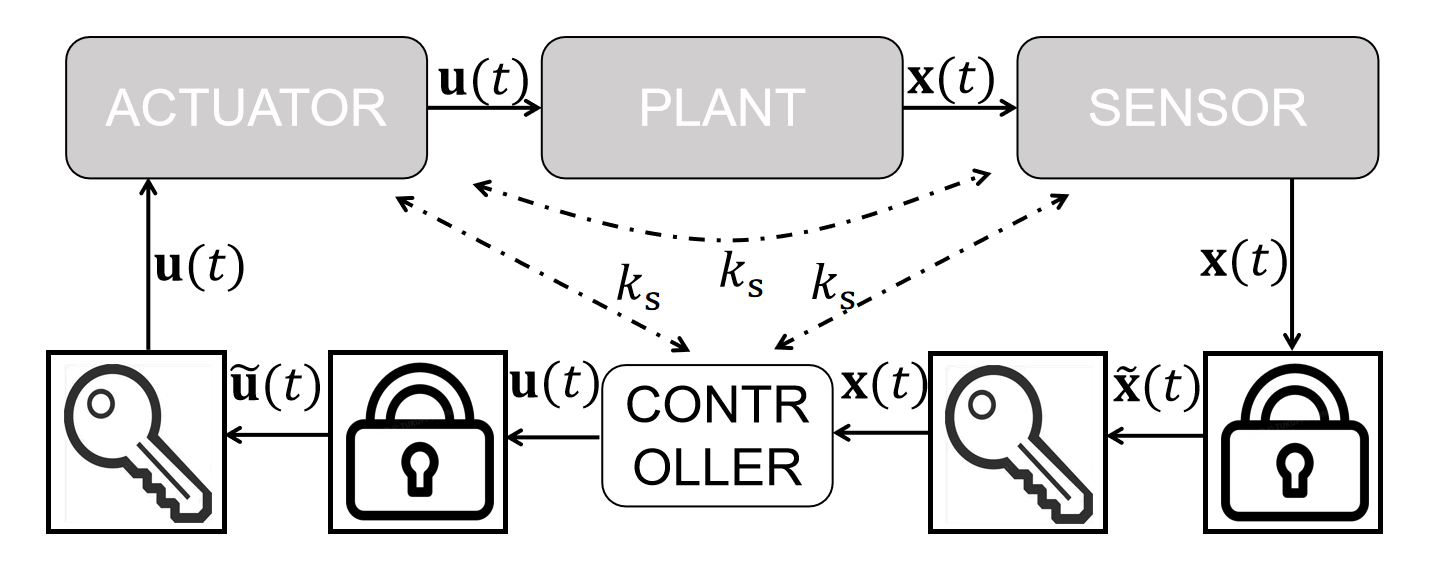}}
	\end{subfigure}
	\centering
	\begin{subfigure}[Asymmetrically Encrypted Control]{
\includegraphics[width=8.5cm]{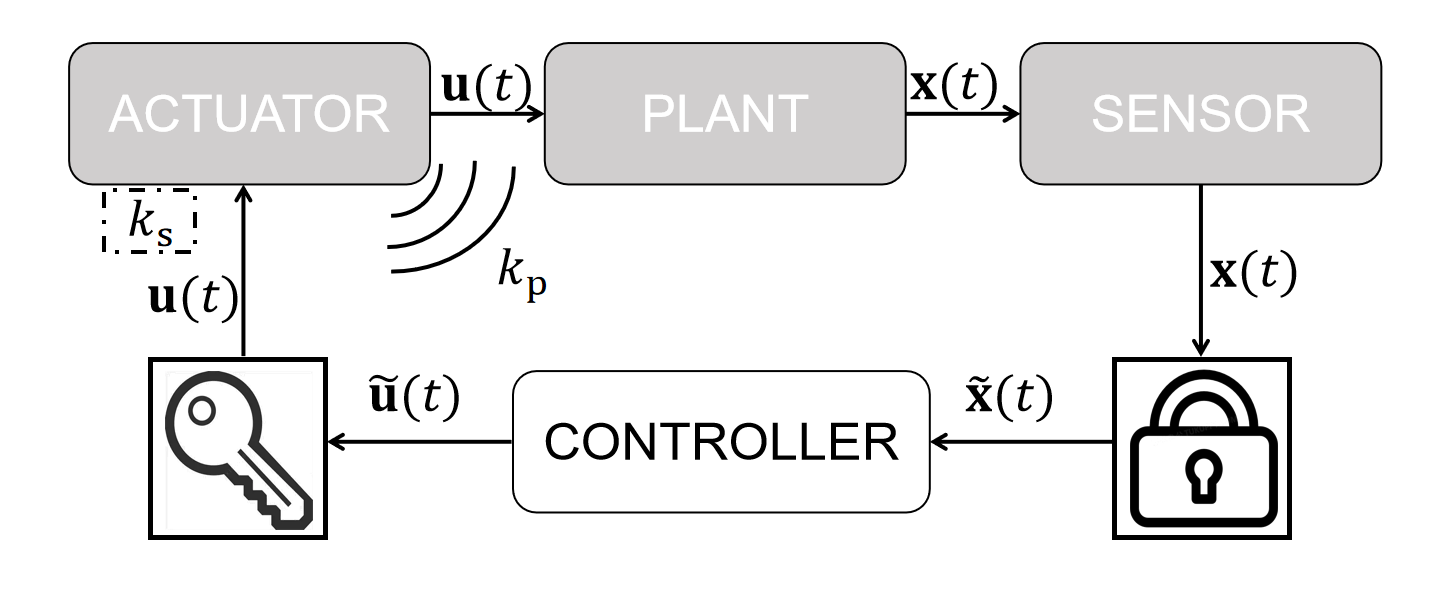}}
	\end{subfigure}
        \caption{Diagrams of classical encrypted control systems.}
    \label{fig.illu_cla}
\end{figure*}

An encrypted control system is a networked control system equipped with encryption schemes that simultaneously ensure control performance and information security, protecting both the plant state and control input from eavesdropping.



\medskip

\noindent{\em Symmetrically Encrypted Control}.  In symmetrically encrypted control \cite{A_AES},  a secret key is first distributed to the actuator, sensor, and controller. In the system, the sensor encrypts the plaintext $\xb$ as $\tilde{\xb}$ and transmits the ciphertext to the controller. The controller decrypts $\tilde{\xb}$ using the secret key, computes the control input
$
\ub = \Kb \xb,
$
re-encrypts it as $\tilde{\ub}$, and sends it to the actuator. The actuator then decrypts $\tilde{\ub}$ and applies it to the plant, completing the control loop.

In this system, as long as the secret key remains undisclosed, the encrypted control system effectively protects against eavesdroppers, including those equipped with quantum computers, on the communication channels between the sensor–controller and controller–actuator. 
However, a critical drawback is that the controller itself possesses the secret key and must decrypt $\tilde{\xb}$. 

\medskip

\noindent{\em Asymmetrically Encrypted Control}. A typical model for asymmetrically encrypted control is studied in \cite{Con_HE1,PAR_CON,con_H,con_H2}. Initially, the actuator generates a key pair and distributes the public key $k_{\rm p}$ to the networked system while securely storing the private key $k_{\rm s}$. Upon obtaining the plant state $\xb$, the sensor encrypts it as $\tilde{\xb}$. The encrypted state $\tilde{\xb}$ is then transmitted to the controller. 
The controller performs computations directly on $\tilde{\xb}$ without decrypting it, and subsequently transmits the result $\tilde{\ub}$ to the actuator, where $\tilde{\ub}$ is decrypted and applied to the plant.  The controller’s ability to operate on the ciphertext $\tilde{\xb}$ is due to the homomorphic property of the encryption scheme, which is a feature of most asymmetric encryption schemes  \cite{RSA,ELE,PAR}. Interested readers may refer to \cite{Con_HE1,PAR_CON} for further details. 

In asymmetrically encrypted control system, even if $\tilde{\xb}$ and $\tilde{\ub}$ are intercepted, the plaintext cannot be recovered without the private key. Moreover, since the controller operates solely on encrypted data due to the homomorphic property, the risks from eavesdropping in controllers are effectively mitigated. However, the security of asymmetric encryption relies on computational complexity, an eavesdropper equipped with a sufficiently powerful quantum computer could potentially recover the plaintext $\xb$ and $\ub$ from the intercepted ciphertexts $\tilde{\xb}$ and $\tilde{\ub}$. The process of these two encrypted systems are illustrated in Fig. \ref{fig.illu_cla}.

\medskip

\subsection{The Confidentiality–Robustness–Efficiency Trilemma}

 \begin{table*}
    \centering
    \caption{Confidentiality, robustness, and efficiency of encrypted control}
    \begin{tabular}{l|cc}
        \toprule
        \textbf{Security Guarantee} & \textbf{Symmetric} & \textbf{Asymmetric} \\
        \midrule
      In-channel confidentiality & \checkmark & \checkmark\\
      In-controller confidentiality & \texttimes & \checkmark \\
        Resistance to quantum computers & \checkmark & \texttimes \\
        \midrule
        Resilience to key errors & Hight & Low \\
        \midrule
        Computation demands & Low & High \\
        \bottomrule
    \end{tabular}
    \label{tab:1}
\end{table*}

Table~\ref{tab:1} summarizes the characteristics of the two encrypted control systems. Symmetric encryption can resist eavesdropping in communication channels but remains vulnerable within controllers because the secret key must be shared with them. However, it is secure against quantum computer attacks, as its protection relies solely on the confidentiality of the key. In contrast, asymmetric encryption safeguards both communication channels and controllers but is susceptible to quantum attacks since its security depends on mathematical problems that quantum computers can efficiently solve \cite{SHOR}.

Moreover, asymmetric encryption entails substantially higher computational complexity than symmetric encryption, as it involves operations such as modular exponentiation or elliptic-curve point multiplication of the key. Symmetric encryption, by comparison, uses lightweight operations such as bitwise XOR and substitution, making it considerably faster to execute.
 In addition, the differences in underlying mathematical operations make asymmetric encryption far more sensitive to key inaccuracies. Even a very small deviation in the key can, due to the nonlinear processes involved, cause the decrypted output to become completely incorrect. By contrast, in symmetric encryption methods, errors in specific portions of the key only affect the corresponding bits of the decrypted message, rather than corrupting the entire output.

In this paper, we aim to address this confidentiality-robustness-efficiency trilemma by  developing a computationally efficient and quantum-resistant encrypted control framework that integrates the strengths of both symmetric and asymmetric encryption. Our purpose relies on quantum entanglements and quantum key sharing between the sensor and actuator.

\section{Quantum Encrypted Control}
\label{sec.qec}

In this section, we present a quantum encrypted control framework. We propose establishing a quantum channel between the sensor and actuator, enabling them to share entangled quantum states.

\subsection{Entangled Sensor and Actuator}

\begin{figure}[t]
    \centering
\includegraphics[width=9cm]{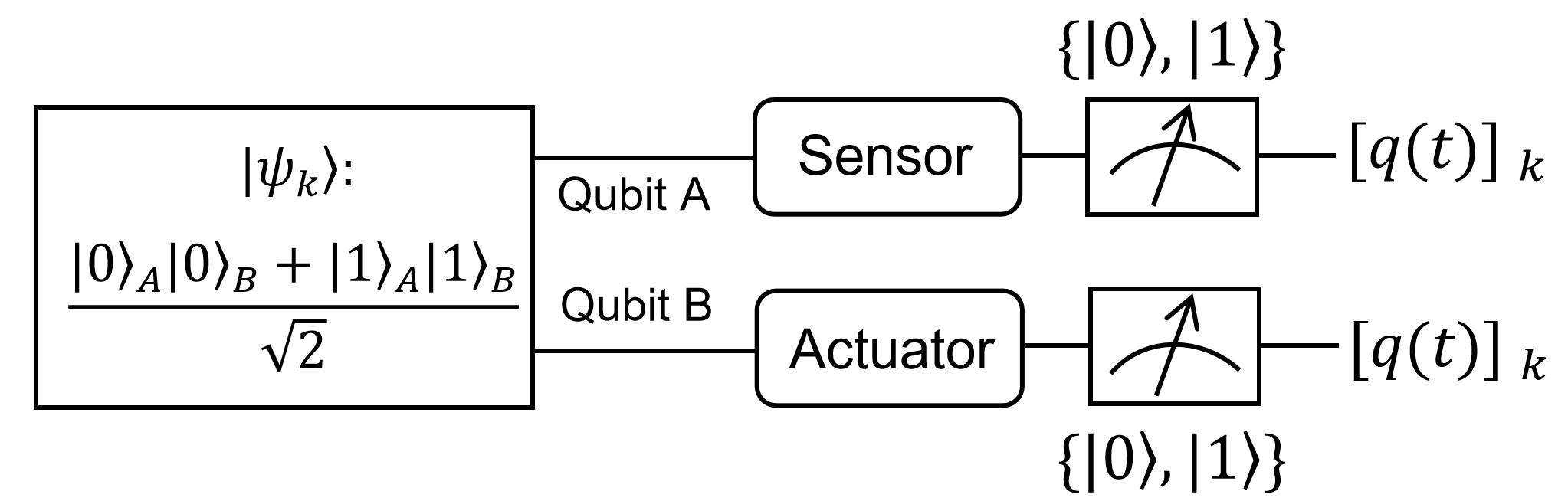}
    \caption{Illustration of the key sharing between  the sensor and actuator through quantum entanglements.
    }
    \label{fig.illu_qun}
\end{figure}

\begin{mybox}{Quantum Channel and Quantum Keys}
\label{box1}
The following standard quantum communication protocol establishes identical quantum keys $q (t) \in\{0,1\}_{w_{\rm q}}$  at the sensor and the actuator over the quantum channel, as illustrated in Fig. \ref{fig.illu_qun}.  
\begin{itemize}
\item[(i)] At each time step $t \in \mathbb{N}$, $w_{\rm q}$ identical Bell states 
\begin{equation}
    \label{eq:per_psi}
|\psi_1^\ast\rangle=\dots= |\psi_{w_{\rm q}}^\ast \rangle=\frac{1}{\sqrt{2}} |0\rangle_A |0\rangle_B + \frac{1}{\sqrt{2}} |1\rangle_A |1\rangle_B 
\end{equation}
are generated through the sensor–actuator quantum channel,  where the first qubit of each pair is held by the sensor and the second by the actuator. 

\item[(ii)] The sensor and the actuator perform measurements on their respective qubits sequentially under the basis $\{|0\rangle,|1\rangle\}$, and record $0$ if the measurement outcome is $|0\rangle$ and $1$ otherwise. 

\item[(iii)] The sensor and actuator collect the sequence of measurement outcomes and obtain a sequence of binary numbers denoted by $q (t) \in\{0,1\}_{w_{\rm q}}$. We term the $q (t)$ quantum keys. 

\end{itemize}
\end{mybox}

Quantum states describe the physical properties of quantum systems, which are ubiquitous in the microscopic world, such as photon polarizations and electron spins. Mathematically, a quantum state is represented by a unit vector in a Hilbert space, denoted $|\psi \rangle$ in Dirac notation. The simplest quantum system is a qubit, which resides in a two-dimensional Hilbert space with orthogonal basis states $|0\rangle$ and $|1\rangle$. Quantum entanglement is a phenomenon in which two or more particles share a correlated quantum state, such that measuring one particle immediately affects the state of the other, regardless of the distance between them. The state space of a composite quantum system is the tensor product of the individual subsystems’ spaces. For example, a pair of qubits A and B can jointly occupy the state
$
|\psi^\ast \rangle =\frac{1}{\sqrt{2}} |0\rangle_A |0\rangle_B + \frac{1}{\sqrt{2}} |1\rangle_A |1\rangle_B,
$
which cannot be factorized into the form $|\psi_1 \rangle_A \otimes |\psi_2 \rangle_B$ and is therefore entangled. This state, one of the Bell states, is widely used in quantum communication protocols \cite{swap}.

Quantum key distribution between the sensor and the actuator is shown in Box ``Quantum Channel and Quantum Keys''. In this process, quantum mechanics ensures that measuring the first qubit of the Bell state $|\psi^\ast\rangle$ in the basis ${|0\rangle,|1\rangle}$ yields a random outcome, $|0\rangle$ or $|1\rangle$, with equal probability. If the outcome $|0\rangle$ is observed, the measured qubit collapses to $|0\rangle$, and a subsequent measurement of the second qubit in the same basis will necessarily yield $|0\rangle$. The same holds if $|1\rangle$ is observed first. Therefore, the results of the measurement on the sensor and actuator are guaranteed to agree according to the laws of quantum mechanics, that is, they share the same key $q(t)$.
In addition, each bit in $q(t)$ is independent and identically distributed (i.i.d.). In practice, such entangled Bell states can be generated using quantum photonic technologies, for example via spontaneous parametric down-conversion or photonic integrated circuits, and distributed over optical channels \cite{Impe}. The security of the quantum key $q(t)$  can be further enhanced using advanced quantum key distribution (QKD) protocols, e.g., \cite{E91}.   QKD relies on fundamental quantum principles: unknown quantum states cannot be perfectly copied and any measurement on a quantum state inevitably disturbs it \cite{swap}. Consequently, secure key exchange can be established without relying on computational assumptions, and any eavesdropping attempt can be detected by the protocol.

\subsection{Quantum Encrypted Control Scheme}
With quantum keys held by the actuator and the sensor, we propose the following quantum encrypted control scheme. We denote 
\[
\ba
&f_{\rm Enc} : \mathbb{R}^n \times \{0,1\}_{w_{\rm q}} \rightarrow \mathbb{C}_{x},\quad  f_{\rm Con} : \mathbb{R}^{1\times n} \times \mathbb{C}_{x} \rightarrow \mathbb{C}_u, \quad f_{\rm Dec} : \mathbb{C}_u \times \{0,1\}_{w_{\rm q}} \rightarrow \mathbb{R}
\ea
\]
as the encryption function, the control law function, and the decryption function, respectively, where $\mathbb{C}_x$ and $\mathbb{C}_u$ denote certain ciphertext spaces of $x$ and $u$, respectively. These functions may vary, as shown in the next subsection. 

\begin{algorithm}[t]
\floatname{algorithm}{Scheme}
\renewcommand{\thealgorithm}{}
\caption{Quantum Encrypted Control ({\sf{QEC})}}
\label{QEC}
\begin{algorithmic}[1]
\State \textbf{Initialization:}
\State Generate quantum encrypted control functions $f_{\rm Enc}$, $f_{\rm Con}$, $f_{\rm Dec}$ and distribute them to the sensor, controller, and actuator, respectively.

\For{$t \in \mathbb{N}$}
    \State \textbf{Quantum Channel:}
    \State \quad Distribute quantum keys $q(t) \in \{0,1\}_{w_{\rm q}}$ to the sensor and the actuator, respectively.
    
    \State \textbf{Sensor:}
    \State \quad 1. Obtain the plant state $\xb(t) \in \mathbb{R}^n$.
    \State \quad 2. Encrypt $\xb(t)$ as $\tilde{\xb}(t) = f_{\rm Enc}(\xb(t), q(t)) \in \mathbb{C}_x$.
    \State \quad 3. Transmit $\tilde{\xb}(t)$ to the controller.

    \State \textbf{Controller:}
    \State \quad 1. Compute $\tilde{\ub}(t) = f_{\rm Con}(\Kb, \tilde{\xb}(t)) \in \mathbb{C}_u$.
    \State \quad 2. Transmit $\tilde{\ub}(t)$ to the actuator.

    \State \textbf{Actuator:}
    \State \quad 1. Decrypt $\tilde{\ub}(t)$ as $\ub(t) = f_{\rm Dec}(\tilde{\ub}(t), q(t)) \in \mathbb{R}$.
    \State \quad 2. Apply $\ub(t)$ to the plant.
\EndFor
\end{algorithmic}
\end{algorithm}

\begin{figure}[t]
\centering
\includegraphics[width=9cm]{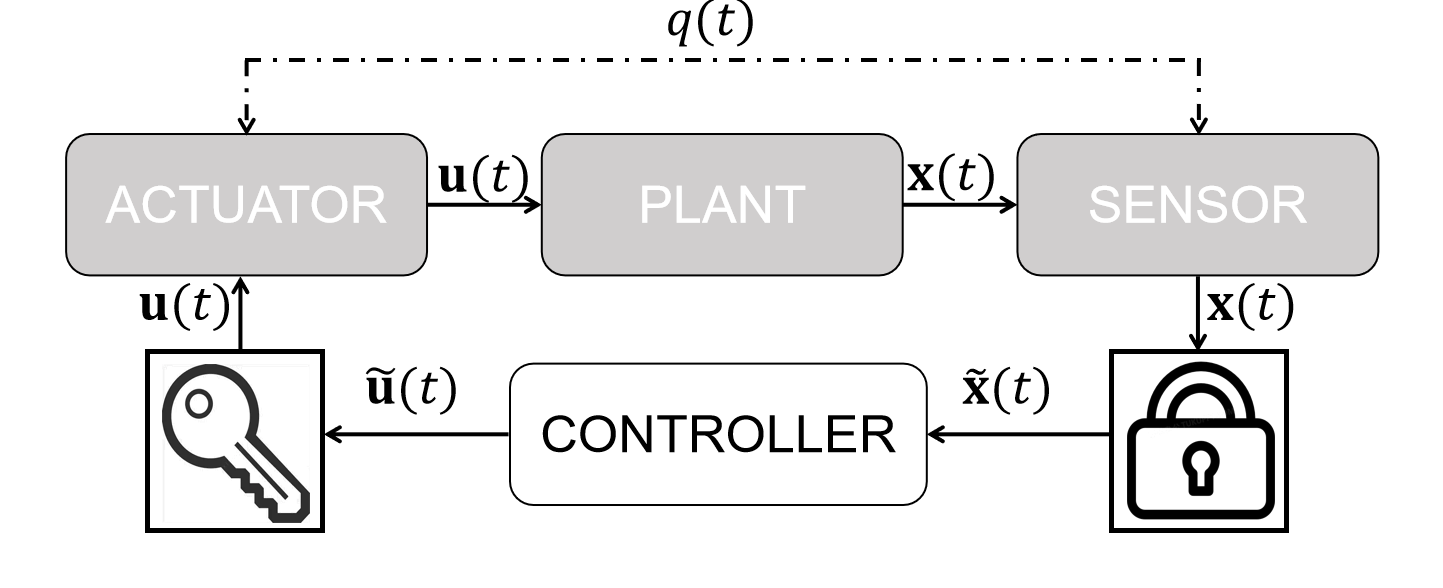}
    \caption{The Diagram of the quantum encrypted control system.}
    \label{fig.illu}
\end{figure}

The proposed scheme $\mathsf{QEC}$ replaces the pair of public–private keys used in classical asymmetrically encrypted control with a pair of identical quantum keys shared between the actuator and the sensor, as shown in Fig. \ref{fig.illu}.
This key replacement enables the design of computationally efficient functions $f_{\rm Enc}$, $f_{\rm Con}$, and $f_{\rm Dec}$, avoiding the heavy computational burden and vulnerability to key errors of classical asymmetric encryption, while achieving a certain level of confidentiality.

\subsection{An Algorithmic Realization}

Next, we construct an algorithm to realize $\mathsf{QEC}$. For simplicity, in the following, we denote  
\[
q(t)=\overline{b_{1,w_{b}-1}b_{1,w_b-2}...b_{1,0}...b_{n,w_b-1}b_{n,w_b-2}...b_{n,0}}(t). 
\]
Our goal is to encrypt each dimension of \(\xb\), \([\xb]_i\) (\(i=1,2,\dots,n\)). To this end, each \(q(t)\) is divided into \(n\) groups of \(w_b\) bits, satisfying \(n w_b = w_{\rm q}\).

\begin{mybox}{Exponential-Logarithmic Realization of  $\mathsf{QEC}$}
From $q(t)$ we define for $i=1,...,n$
\begin{equation}
\label{eq:beta}
\ba
    \beta_i(t): = -(2^{w_b-1}+1){b_{i,w_b-1}}(t)+\sum_{j=0}^{w_{b}-2}2^{j}b_{i,j}(t)+1.
    \ea
\end{equation}

\begin{itemize}
    \item[(i)] {\bf Encryption function}: The function $f_{\rm Enc}$ under which $\tilde{\xb}(t)=f_{\rm Enc}(\xb(t), q(t))$ is defined by
\begin{equation}
    \label{eq:enc}
    \ba
    \ [\tilde{\xb}]_i &= \exp({[\xb]_i / \beta_i(t)}), \ i=1,2,...,n. 
    \ea
\end{equation}

\item [(ii)] {\bf Control law function}: The function $f_{\rm Con}$  under which $\tilde{\ub}(t)=f_{\rm Con}(\Kb, \tilde{\xb}(t))$ is defined by 
\begin{equation}
    \label{eq:con}
    [\tilde{\ub}]_i = [\tilde{\xb}]_i^{[\Kb]_{i}}, \ i=1,2,...,n.
\end{equation}

\item [(iii)] {\bf Decryption function}: The function $f_{\rm Dec}$ under which $\ub(t)= f_{\rm Dec}(\tilde{\ub}(t), q(t))$ is defined by
\begin{equation}
    \label{eq:dec}
    \ba
   \ub &= \sum_{i=1}^n{\beta}_i(t) \ln([\tilde{\ub}]_i).
   \ea
\end{equation}

\end{itemize}
\end{mybox}


The proposed Exponential-Logarithmic Realization is given by \eqref{eq:beta}-\eqref{eq:dec}. It  involves only exponential, logarithmic, and linear operations, which are suitable for high-speed processing and for high-dimensional systems. This contrasts with modular exponentiation and elliptic-curve point multiplication, which are used in asymmetric encryption. Despite its simplicity, the overall scheme offers the following security guarantees:
\begin{itemize}
\item [(i)]  {\bf Resistance to known-ciphertext attacks (KCA)} describes that the probability of an eavesdropper successfully recovering the corresponding plaintext from a given ciphertext is minimal. For this security, note that each bit in $q(t)$ has equal probability to be $0$ or $1$, it can be verified that the coefficients ${\beta}_i(t)$ are nonzero integers uniformly distributed in $\{-2^{w_b-1},-2^{w_b-1}+1,...,-1,1,..., 2^{w_b-1}\}$. Therefore, for a given ciphertext $[\tilde{\xb}(t)]_i$ with $1\leq i\leq n$ and $t\in\mathbb{N}$, its corresponding plaintext $[\xb(t)]_i=\beta_i(t)\ln([\tilde{\xb}(t)]_i)$ is also uniformly distributed.  Then, an adversary can correctly guess the plaintext only with probability $\frac{1}{2^{w_b}}$, which will be arbitrarily small if the number of transmitted Bell states $w_{\rm q}$ can be large enough.

\item[(ii)] {\bf Resistance to known-plaintext attacks (KPA)} is another fundamental concept in cryptography, describing a cipher’s ability to remain secure even if an attacker has access to some pairs of plaintexts and their corresponding ciphertexts.  To investigate KPA, we note that the ciphertexts of $[\xb(t)]_i$ are generated using distinct keys $\beta_i(t)$ across both time steps $t\in\mathbb{N}$ and dimensions $i=1,2,...,n$.
Since each bit in $q(t)$ is  i.i.d., we know $\beta_i(t)$ is also i.i.d. with respect to both $t$ and $i$. Then, the exposure of a subset of some plaintext-ciphertext pairs together with their associated keys doesn't enable an eavesdropper to infer any of the remaining keys, nor decrypt any other ciphertexts.


\end{itemize}
The effectiveness of the Exponential-Logarithmic Realization for {\sf{QEC}} is shown in the following result. 

\begin{theorem}
\label{thm:1}
The state \(\xb(t)\) and control input \(\ub(t)\) generated   by \eqref{eq:enc}-\eqref{eq:dec}  satisfy
\[
    \ub(t) = \Kb \xb(t), \quad \forall t \in \mathbb{N}.
\]
\end{theorem}
{\em Proof}. 
For all $t\in\mathbb{N}$,
by \eqref{eq:dec} and \eqref{eq:con},  we have
\[
\ba
\ub(t)=&\sum_{i=1}^n\hbeta_i(t)\ln([\tilde\ub(t)]_i)
=\sum_{i=1}^n\hbeta_i(t)[\Kb]_{i}\ln([\tilde{\xb}(t)]_i),
\ea\] 
Then from \eqref{eq:enc} we obtain
\[\ba
\ \ub(t)=\sum_{i=1}^n\frac{\beta_i(t)[\Kb]_{i}[\xb(t)]_i}{\beta_i(t)}=\Kb\xb(t).
\ea
\]
The proof is complete. \hfill$\square$

\section{Stability under Intrinsic Quantum Error}
\label{sec.mat}
 
 In practice, the generated Bell state may have an intrinsic quantum error arising from errors introduced during the preparation of the quantum state and noise encountered during its transmission, e.g., \cite{pre_er}. We impose the following assumption to describe a more general model where each Bell state takes the imperfect form. For simplicity, we assume all imperfect states take the \emph{same} form.

\medskip

\begin{assumption}[Quantum Key Errors]
\label{ass2}
At each time step $t \in \mathbb{N}$, each Bell state distributed between the sensor and the actuator takes the form 
$$|\psi\rangle=a|0\rangle_A |0\rangle_B + b|0\rangle_A |1\rangle_B + c|1\rangle_A |0\rangle_B + d|1\rangle_A |1\rangle_B.
$$
for some $a,b,c,d\in \mathbb{C}$. 
\end{assumption}

\medskip

By \cite{swap}, there holds $|a|^2+|b|^2+|c|^2+|d|^2=1$. 
Then it can be noticed that the perfect Bell state in \eqref{eq:per_psi} can be obtained by letting $a=d$ and $b=c=0$. For the case $b,c\neq0$, the terms $b|0\rangle_A |1\rangle_B$  and $c|1\rangle_A |0\rangle_B$ capture uncertainties in the Bell states, which will lead to mismatches between the keys at the sensor and actuator. As a result, we denote the obtained keys in the sensor and the actuator as
\begin{equation}
    \label{eq:keys}
\ba
\qse(t)=\overline{b_{1,w_{b}-1}b_{1,w_b-2}...b_{1,0}...b_{n,w_b-1}b_{n,w_b-2}...b_{n,0}}(t)\\
\qac(t)=\overline{\hat{b}_{1,w_{b}-1}\hat{b}_{1,w_b-2}...\hat{b}_{1,0}...\hat{b}_{n,w_b-1}\hat{b}_{n,w_b-2}...\hat{b}_{n,0}}(t)\\
\ea
\end{equation}
Thus, we present the following lemma to illustrate the mismatch between  $\qse(t)$ and $\qac(t)$.
\begin{lemma}
\label{propo}
 Under Assumption \ref{ass2}, independent of  $t$, for any fixed $t\in\mathbb{N}$, the keys $\qse(t),\qac(t)$ defined in \eqref{eq:keys} satisfy
\begin{itemize}
    \item [(i)]Each bit $b_{i,j}(t)$ is i.i.d. for $i=1,2,...,n$ and $j = 0, 1, \dots, w_{b}-1$.
\item[(ii)] Each bit $\hat{b}_{i,j}(t)$ satisfies
\[
\ba
\mathbb{P}\big(\hat{b}_{i,j}(t) \neq B \,\big|\, b_{i,j}(t)=B )=p,\\ B=0,1,\ \ \  i=1,2,...,n, \\  j = 0, 1, \dots, w_{b}-1,
\ea
\]
for \( p =|b|^2+|c|^2 \), where the error events \(\{\hat{b}_{i,j}(t) \neq b_{i,j}(t)\}\) are i.i.d. across all bit positions and independent of \(\{b_{i,j}\}\).
\end{itemize}
\end{lemma}
{\em Proof}. The proof of Lemma \ref{propo} follows fundamental principles of quantum mechanics \cite{swap}. \hfill$\square$

The following theorem shows that the closed-loop system \eqref{eq:control}-\eqref{eq:controllaw} encrypted with $\mathsf{QEC}$ remains stable as long as the key error probability \(p\) is sufficiently small.

\begin{theorem}
\label{thm:3}
Suppose Assumption  \ref{ass2} holds and that the ideal system \eqref{eq:control}-\eqref{eq:controllaw} is closed-loop stable, i.e., $\rho(\Ab+\Bb\Kb)<1$. Then, there exists a positive value given by\[
\ba
p \leq p^\ast =& \frac{\epsilon}{\max\{16\|\Mb_1\|^2_\ast,16\|\Mb_1\|_\ast,4A\|\Mb_1\|^2_\ast\}},
\ea
\]
where $\epsilon=1-\rho(\Ab+\Bb\Kb)$, $A:=\frac{4^{w_b} + 3 \cdot 2^{w_b} + 2}{3\mathsf{E}(\beta_i^2(t))}$, and $\|\cdot\|_\ast$ is some induced matrix norm determined by $\Ab+\Bb\Kb$, such that for all $|b|^2+|c|^2\leq p^\ast$,
there holds $$
\lim_{t\to\infty} \mathsf{E}(\xb(t)\xb^\top(t))=0
$$
along the Exponential-Logarithmic Realization  $\mathsf{QEC}$ \eqref{eq:enc}-\eqref{eq:dec}. 
\end{theorem}

The proof of Theorem \ref{thm:3} can be seen in Appendix \ref{app:thm2}.

\section{Quantized Implementation of {\sf{QEC}}}
\label{sec.quan}

In practical implementations, the communication links between the sensor and controller, and between the controller and actuator, have limited bandwidth. Consequently, the ciphertexts $\tilde{\xb}$ and $\tilde{\ub}$ must be represented in binary form.  
Let the bandwidths of the sensor–controller and controller–actuator communication channels be $nw$ and $w$ bits, respectively, for some $w \in \mathbb{N}_+$. This implies that each element of $\tilde{\xb}$ and $\tilde{\ub}$ is represented using $w$ bits. Hence, the ciphertext space is given by $\mathbb{C} = \{0,1\}^{w}$.

\subsection{{\sf{QEC}} with Quantization}

To formalize the quantization process, we define the following stochastic quantization function.

\begin{algorithm}[H]
\floatname{algorithm}{Function}
\renewcommand{\thealgorithm}{}
\caption{Quantization Function $Q_w:\mathbb{R}\rightarrow\{0,1\}_{w}$}
\label{QEF1}
\begin{algorithmic}[1]
\State \textbf{Input:} $v \in \mathbb{R}$
\State \textbf{Initialization:} Prepare $w$ empty bits $\overline{a_{w-1}a_{w-2}\ldots a_{0}}$.
\State Define
\begin{equation}
    \label{eq:vfun}
y=g(v) =
\begin{cases}
v, & v > 1 \\[2mm]
2 - \dfrac{1}{v}, & v \le 1
\end{cases}
\end{equation}
\State Let $\overline{a_{w-1}a_{w-2}\ldots a_{1}a_0}=h_w(y)$, where the stochastic function  $h_w(y):\mathbb{R}\to\{0,1\}_w$ is defined by 
\begin{equation}
    \label{eq:hde}
\ba
&\mathbb{P}\left(\sum_{j=0}^{w-1} 2^{-j} a_{j}=\frac{\lfloor 2^{w-1}y\rfloor}{2^{w-1}}\right) = 1-(2^{w-1}y-\lfloor 2^{w-1}y\rfloor),\\
&\mathbb{P}\left(\sum_{j=0}^{w-1} 2^{-j} a_{j}=\frac{\lfloor 2^{w-1}y\rfloor}{2^{w-1}}+\frac{1}{2^{w-1}}\right) = 2^{w-1}y-\lfloor 2^{w-1}y\rfloor;
\ea
\end{equation}

\State \textbf{Output:} $Q_w(x) = \overline{a_{w-1}a_{w-2}\ldots a_{0}}$.
\end{algorithmic}
\end{algorithm}

For such a stochastic quantization function $\overline{a_{w-1}a_{w-2}...a_0}=Q_w(v)$, we can directly obtain that, for all $  v\in \left[\frac{1}{2},2\right]$, it hold 
\begin{equation}
    \label{eq:qw0}
    \ba
\mathsf{E}\left(\sum_{j=0}^{w-1}2^{-j}a_{j}-g(v)\right)&= 0,   \\ 
    \ea
\end{equation}
\begin{equation}
    \label{eq:qw}
    \ba
\mathsf{E}\left(\sum_{j=0}^{w-1}2^{-j}a_{j}-g(v)\right)^2&\leq \frac{1}{2^{2w}}.   \ 
    \ea
\end{equation}

\begin{mybox}{Exponential-Logarithm Realization of  $\mathsf{QEC}$ with Quantization}
The parameters $\alpha,\overline{x}_i,\delta_i>0$ for $i=1,2,...,n$ are globally shared. From $q(t)$, we define
\begin{equation}
    \label{eq:beta_q}
\ba
    \beta_i(t): =& -(2^{w_b-1}+1){b_{i,w_b-1}}(t)+\sum_{j=0}^{w_{b}-2}2^{j}b_{i,j}(t)+1+(-1)^{b_{i,w_{b-1}}(t)}\frac{\overline{x}_i}{\ln (1+\alpha)},
    \ea
\end{equation}
for $i=1,2,...,n$.

\begin{itemize}
    \item [(i)] {\bf Encryption function}: The function $f_{\rm Enc}$ under which  $\tilde{\xb}(t)=f_{\rm Enc}(\xb(t), q(t))$ is defined by 
\begin{equation}
    \label{eq:enc2}
    \ba
    v_i(t) &= \exp({[\xb(t)]_i / \beta_i(t)}),  \\
[\tilde{\xb}]_i(t) &=\overline{c_{i,w-1}c_{i,w-2}\ldots c_{i,0}}(t)= Q_{w}(v_i(t)), 
    \ea
    \end{equation}
for $i=1,2,...,n$.
    
    \vspace{0.3em}

    \item [(ii)] \textbf{Control law function:}  
The function $f_{\rm Con}$ under which  $\tilde{\ub}(t)=f_{\rm Con}(\Kb, \tilde{\xb}(t))$ is defined by
    \begin{equation}
        \label{eq:con2}
        \begin{aligned}
            z_i &=  
            \left(
                g^{-1}\left(\sum_{j=0}^{w-1} 2^{-j} c_{i,j} \right)
            \right)^{ [\Kb]_i}, \\
            [\tilde{\ub}]_i &= \overline{d_{i,w-1}d_{i,w-2}\ldots d_{i,0}}= Q_{w}\left(z_i^{1/\delta_i}\right), 
        \end{aligned}
    \end{equation}
for $i=1,2,...,n$.

    \item [(iii)] {\bf Decryption function}: The function $f_{\rm Dec}$ under which $\ub(t)= f_{\rm Dec}(\tilde{\ub}(t), q(t))$ is defined by 
\begin{equation}
    \label{eq:dec2}
    \ba
\ub&=\sum_{i=1}^n\delta_i{\beta}_i \ln\left(g^{-1}\left(\sum_{j=0}^{w-1} 2^{-j} d_{i,j}\right)\right).
    \ea
\end{equation}

\end{itemize}





\end{mybox}

The stochastic quantizer proposed in this work is inspired by~\cite{dithered}. In that work, the quantizer maps each value to its two nearest quantization levels in a probabilistic manner. The proposed quantizer $Q_w({\cdot})$ follows the same quantization principle. However, the key difference is that $Q_w(\cdot)$ is defined over $w$-bit binary numbers, $\{0,1\}_w$, making it directly applicable in bit-constrained communication scenarios. Moreover, the function $g(\cdot)$ is initially applied to the input $v$. This is because the ciphertext  to be quantized in $\mathsf{QEC}$ has an exponential form, its values within the interval $\left[\frac{1}{2},2\right]$ exhibit symmetry around $1$. Specifically, for $e^{-x} \in \left[\frac{1}{2},1\right]$, the value of $e^x$ lies within $\left[1,2\right]$. To exploit this symmetry, $g(v)$ in~\eqref{eq:vfun} leaves values of $v>1$ unchanged and maps $v \in \left[\frac{1}{2},1\right]$ to the range $\left[0,1\right]$. Subsequently, after quantizing $y=g(v) \in \left[0,2\right]$ using \eqref{eq:hde}, a unified quantization error, as described in \eqref{eq:qw0}-\eqref{eq:qw}, can be derived for $v \in \left[\frac{1}{2},2\right]$.

The use of stochastic quantization is motivated by its following advantages. Stochastic quantization, known as dithering in the signal processing literature, has been widely used to suppress undesirable nonlinear effects such as limit cycles in quantized systems by decorrelating the quantization error from the signal, e.g., \cite{Gray1993,book2008}. In the control literature,  a fundamental result in \cite{Delchamps1990} established that for discrete-time linear time-invariant systems under deterministic quantization,  no control strategy can stabilize the system in the closed-loop in the sense that all trajectories converge asymptotically to zero.


Now, we are ready to present the following algorithmic realization of the $\mathsf{QEC}$ by applying quantization into \eqref{eq:enc}-\eqref{eq:dec}, as shown in \eqref{eq:beta_q}-\eqref{eq:dec2}. 

\subsection{Effect of Quantization Error}
Since $\tilde{\xb},\tilde{\ub} \in \{0,1\}^{nw}$, 
both the plant state and control input can be transmitted through limited-bandwidth communication channels. 
On the other hand, the finite bandwidth inevitably introduces quantization errors. 
The following theorem characterizes the error bound induced by communication quantization.
\begin{theorem}
\label{thm:2}
For the state \(\xb(t)\) and control input \(\ub(t)\) generated by \eqref{eq:enc2}-\eqref{eq:dec2} with $\alpha\in(0,1]$, 
$
\delta_i \geq |[\Kb]_i|
$,  $\overline{x}_i>{|[\xb(t)]_i|}$, for $i=1,2,...,n$ and all $t\in\mathbb{N}$, the following holds:
\[
    \mathsf{E}\left( \ub(t) - \Kb \xb(t) \right)^2
    \leq 
    {\sum_{i=1}^n } 
    \frac{5(\beta_i(t) \delta_i)^2}{2^{2w}}
    + o\!\left(\frac{1}{2^{2w}}\right),
    \quad \forall\, t \in \mathbb{N}.
\]
\end{theorem}
The proof of Theorem \ref{thm:2} can be seen in Appendix \ref{app:thm3}.

\section{Quantum-Classical Feedback in Confidentiality}\label{DP}

Although the introduction of quantization has introduced errors, it simultaneously provides  {\em privacy protection}. Quantization helps privacy protection by converting continuous data streams into discrete values, thereby limiting information leakage.
Regarding this point, existing literature has extensively investigated the privacy protection of system parameters or initial states through quantization \cite{Murguia2018,Kawano2021,Liu2025Design}.

In this section,  we discuss how quantization provides privacy protection to quantum keys under the notion of differential privacy, completing the feedback loop of confidentiality between the control system signals and quantum keys. 

\subsection{Differential Privacy of Quantum Keys}

We introduce the following quantum key matrix\footnote{Note that the $\beta_i(t),i=1,\dots,n$ are equivalent to the quantum key $q(t)$.  } associated with the time interval $t \in [0, T]$
\[
\mathbf{E}_T := \begin{bmatrix}
    \beta_1(0) & \beta_2(0) & \dots & \beta_n(0) \\
    \beta_1(1) & \beta_2(1) & \dots & \beta_n(1) \\
    \vdots & \vdots & \ddots & \vdots \\
    \beta_1(T-1) & \beta_2(T-1) & \dots & \beta_n(T-1)
\end{bmatrix}.
\]
It is important to note that all the information $ \tilde{\mathbf{x}}^\top(t)$, $\tilde{\mathbf{u}}^\top(t) $ sent through the classical communication channel over the time interval $t \in [0, T]$ can be viewed as the image of a function $O(\cdot): \mathbb{R}^{n \times T} \to \{0, 1\}_w^{2n \times T}$, where
\[
O_T(\mathbf{E}_T) := \begin{bmatrix}
    \tilde{\mathbf{x}}^\top(0) & \tilde{\mathbf{u}}^\top(0) \\
    \tilde{\mathbf{x}}^\top(1) & \tilde{\mathbf{u}}^\top(1) \\
    \vdots & \vdots \\
    \tilde{\mathbf{x}}^\top(T-1) & \tilde{\mathbf{u}}^\top(T-1)
\end{bmatrix}.
\]

To define privacy in this context, we introduce the following $\zeta$-adjacency. \begin{definition}
   Given $\zeta \geq 0$ and a pair of   $\mathbf{E}_T, \mathbf{E}_T' \in \mathbb{R}^{n \times T}$ generated by quantum keys, where the elements in $\mathbf{E}_T, \mathbf{E}_T'$ are denoted as $\beta_i(t), \beta_i'(t)$ for $i = 1, 2, \dots, n$ and $t = 0, 1, \dots, T-1$, respectively. Then, $\mathbf{E}_T, \mathbf{E}_T' \in \mathbb{R}^{n \times T}$ is said to hold $\zeta$-adjacency, denoted $(\mathbf{E}_T, \mathbf{E}_T') \in \mathrm{Adj}^\zeta$, if for some $i^\ast \in \{1, 2, \dots, n\}$ and some $t^\ast \in \{0, 1, \dots, T-1\}$, there hold
   $
   \beta_i(t) = \beta_i'(t) \text{ for all } t \neq t^\ast \text{ and } i \neq i^\ast,
   $
   and 
   \begin{equation}
   \label{eq:beta_d}
   \bigg|\, |\beta_i(t) - \beta_i'(t)| - 2^{w_b}\left(1 - \mathrm{sgn}(\beta_i(t)\beta_i'(t))\right) \,\bigg| \le \zeta
   \end{equation}
   for $i = i^\ast$ and $t = t^\ast$.
\end{definition}

    The definition of the $\zeta$-adjacency   formalizes the notion that two quantum key matrices $\mathbf{E}_T$ and $\mathbf{E}_T'$ are ``neighbors'' if they differ in at most one element, located at a specific index $i^*$ and time step $t^*$. And the core of this definition  lies in the condition \eqref{eq:beta_d} governing this differing element, which implements a  distance on the domain 
    \[\ba\beta_i(t)\in\mathcal{A}_\beta:=&\left[-2^{w_b-1}-\frac{\overline{x}_i}{\ln(1+\alpha)},-\frac{\overline{x}_i}{\ln(1+\alpha)}\right)\cup\left(\frac{\overline{x}_i}{\ln(1+\alpha)}, 2^{w_b-1}+\frac{\overline{x}_i}{\ln(1+\alpha)}\right]
    \ea
    \] from \eqref{eq:beta_q}.
This can be intuitively understood by considering two cases for the values $\beta_{i^*}(t^*)$ and $\beta'_{i^*}(t^*)$. First, when they have the same sign, the condition becomes $|\beta_{i^*}(t^*) - \beta_{i^*}'(t^*)| \le \zeta$, which is a standard absolute difference. Second, when they have opposite signs, the condition becomes $\left|\, |\beta_{i^*}(t^*) - \beta_{i^*}'(t^*)| - 2^{w_b} \,\right| \le \zeta$, which means the two values must lie near the opposite ends of the entire domain. Therefore, this definition effectively reduces the sensitive domain of $\mathcal{A}_\beta$ from the four endpoints of its two constituent intervals to just the two endpoints of the entire range, i.e., $-\frac{\overline{x}_i}{\ln(1+\alpha)}$ and  $\frac{\overline{x}_i}{\ln(1+\alpha)}$.

We are now ready to define differential privacy for  $\mathsf{QEC}$. \begin{definition}
    Let $(\{0,1\}_w^{2n\times k}, \mathcal{F}, \mathbb{P})$ be a probability space. For some $k \in \mathbb{N}$, and for some $\epsilon \geq 0$ and $\Delta \geq 0$, $\mathsf{QEC}$ is said to be $(\epsilon, \Delta)$-differentially private under the $\zeta$-adjacency relation at a finite time $T$ if it holds
    \[
    \mathbb{P}(O_T(\mathbf{E}_T) \in \mathbb{O}) \leq e^{\epsilon} \mathbb{P}(O_T(\mathbf{E}_T') \in \mathbb{O}) + \Delta
    \]
    for all $(\mathbf{E}_T, \mathbf{E}_T') \in \mathrm{Adj}^\zeta$ and all $\mathbb{O} \in \mathcal{F}$.
\end{definition}
 This definition ensures that the output of the observation function is not significantly affected by small changes in the quantum key sequences, thus guaranteeing privacy. Building on these definitions, we have the following theorem.
\begin{theorem}
\label{thm:4}
  If $\zeta \frac{(1+\alpha)\ln^2(1+\alpha)}{\overline{x}_i} \in [0, \frac{1}{2^{w-1}}]$, then $\mathsf{QEC}$ realized by \eqref{eq:beta_q}-\eqref{eq:dec2} is $(0, \Delta)$-differentially private under the $\zeta$-adjacency relation at any finite time $T$, where 
  \[
  \Delta = \zeta \frac{(1+\alpha)\ln^2(1+\alpha)}{\overline{x}} 2^{w-1},
  \]
 for  $\overline{x}:=\max_{i=1,2,...,n}\overline{x}_i$.
\end{theorem}

The proof of Theorem \ref{thm:4} can be seen in Appendix \ref{app:thm4}.
This result reveals that the proposed stochastic quantization preserves differential privacy of quantum keys. Remarkably, this establishes a two-way privacy-preserving feedback mechanism between the quantum and classical information processing in the networked system: the quantum keys encrypt system states and control inputs to withstand information-recovery attacks, while the quantization of information over classical channels, in turn, provides differential privacy protection for the quantum keys themselves.   

\subsection{Privacy vs Quantization Error Trade-off}
From Theorems \ref{thm:2} and \ref{thm:4}, a trade-off between the privacy protection and quantization error can be established. We present  the following corollary.
\begin{corollary}
\label{cor:1}
    For given $E_g,\Delta_g,\zeta_g>0$, if \[
w \ge \max\left\{
w_b + 1 + \log_2\left(\frac{\sqrt{5n}\,{\delta}}{E_g}\right),\;
\log_2\left(\frac{20n \overline{x} {\delta}^2 \Delta_g}{E_g^2\zeta_g}
\right)\right\},
\]then parameter $\alpha$ can be chosen as
\[
\alpha\in\left[\frac{2\sqrt{5n}\overline{x}{\delta}}{2^w E_g},e^{\min\left\{\overline{x} / 2^{w_b - 1},{\sqrt{\Delta_g\overline{x} / (2^w \zeta_g)}}\right\}} - 1\right].
\]
In this case, $\mathsf{QEC}$ realized by \eqref{eq:beta_q}-\eqref{eq:dec2} satisfy that 
    \begin{itemize}
        \item [(i)] It is $(0, \Delta_g)$-differentially private under the $\zeta_g$-adjacency relation at any finite time $T$,
        \item [(ii)] The generated state \(\xb(t)\) and control input \(\ub(t)\) satisfy
       \[
    \mathsf{E}\left( \ub(t) - \Kb \xb(t) \right)^2
    \leq 
    E^2_g,\quad t \in \mathbb{N}.
\]
    \end{itemize}
\end{corollary}

Corollary \ref{cor:1} explicitly reveals the trade-off between the quantization error $E_g$ and the privacy level, characterized by $\frac{\Delta_g}{\zeta_g}$. Specifically, with a sufficiently large communication bandwidth $w$, increasing $\alpha$ can effectively reduce $E_g$, but at the cost of a higher $\frac{\Delta_g}{\zeta_g}$; conversely, decreasing $\alpha$ improves privacy protection level by reducing $\frac{\Delta_g}{\zeta_g}$ while incurring a larger $E_g$. The proof of Corollary \ref{cor:1} can be found in Appendix \ref{app:cor1}.

\section{Numerical Simulations}
\label{sec.num}

In this section, we present numerical simulations to validate the theoretical results established in this paper.  

\subsection{Simulation Settings}

We consider a robotic arm control task whose dynamics are linearized around an operating point. The system is characterized by three state variables $\xb = \begin{bmatrix}
    x_1& x_2& x_3
\end{bmatrix}^\top$, representing the joint angular position $x_1$, the joint angular velocity $x_2$, and the motor current $x_3$. The control input $\ub$ corresponds to the applied motor voltage. The continuous-time dynamics of the system are modeled as
\[
\ba
\dot{\xb}(t) 
&=
\begin{bmatrix}
0 & 1 & 0 \\
0 & -\frac{b}{J} & \frac{K_t}{J} \\
0 & -\frac{K_e}{L} & -\frac{R}{L}
\end{bmatrix} \xb(t)
+
\begin{bmatrix}
0 \\
0 \\
\frac{1}{L}
\end{bmatrix} \ub(t),
\ea
\]
where \(J\) denotes the moment of inertia, \(b\) is the viscous friction coefficient, \(K_t\) is the motor torque constant, \(K_e\) is the back electromotive force constant, \(R\) is the armature resistance, and \(L\) is the armature inductance.

For simulation purposes, the parameters are assigned the following numerical values:  
$J = 1$, $b = -2$, $K_t = 1$, $K_e = 1$, $R = 1$, and $L = 0.5$.  
The system is then discretized with a sampling period of $\Delta T = 0.01\, \text{s}$, resulting in the following discrete-time representation.  
\begin{equation}
\label{eq:example_opendloop}
\ba
\xb(t+1)
&=
\begin{bmatrix}
1.0 & 0.1 & 0.0 \\
0.0 & 1.1 & 0.1 \\
0.0 & -0.2 & 0.8
\end{bmatrix}
\xb(t)
+
\begin{bmatrix}
0.001 \\
0.02 \\
0.2
\end{bmatrix}\ub(t).
\ea
\end{equation}

The open-loop system described by \eqref{eq:example_opendloop} is unstable, as indicated by an eigenvalue $\lambda = 1.15>1$. To stabilize the system, we design the following state feedback control law.  
\begin{equation}
\label{eq:example_closedloop}
\ba
\ub(t)
= \Kb \xb(t)=
\begin{bmatrix}
-63 & -25 & 0.78
\end{bmatrix}\xb(t),
\ea
\end{equation}  
which ensures the stability of the closed-loop system.

\subsection{Control Performance of $\mathsf{QEC}$ with Quantum Key Errors}

\begin{figure}[t]
 \centering
\includegraphics[width=8.5cm]{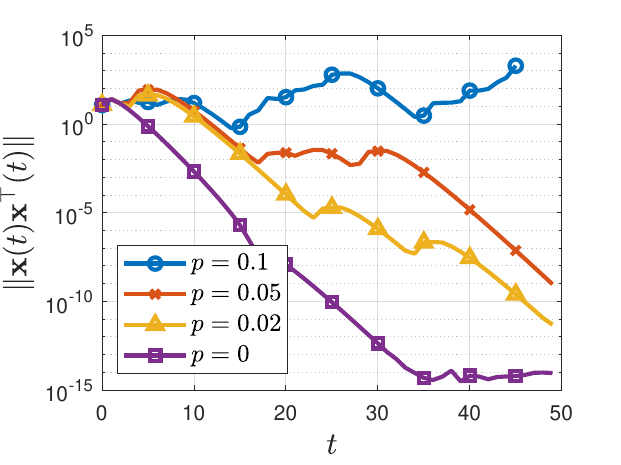}
    \caption{Time evolutions   of $\|\ub(t)-\Kb\xb(t)\|^2$ $\mathsf{QEC}$ in \eqref{eq:example_opendloop}-\eqref{eq:example_closedloop} encrypted by $\mathsf{QEC}$ \eqref{eq:beta}-\eqref{eq:dec} under different quantum error levels $p$.}
    \label{fig.sim1}
\end{figure}

We apply Exponential-Logarithmic Realization $\mathsf{QEC}$ \eqref{eq:beta}-\eqref{eq:dec} under Assumption~\ref{ass2} with different measurement inaccuracy probabilities $p=|b|^2+|c|^2$. According to Theorem \ref{thm:3}, we have a threshold of $p$ as $p^\ast\approx 0.025$. Fig.~\ref{fig.sim1} presents the system response for different values of $p$. The results indicate that when $p\leq p^\ast$, the system state $\xb(t)$ still converges to the equilibrium point without bias, while excessively large measurement inaccuracy leads to instability of the system, verifying Theorem~\ref{thm:3}.

\subsection{Control Performance of $\mathsf{QEC}$ with Quantization}

We consider the case where the encrypted control system is affected by communication quantization. In this case, we apply Exponential-Logarithm Realization $\mathsf{QEC}$ with Quantization \eqref{eq:beta_q}-\eqref{eq:dec2} with different bandwidth $w$ and fixed parameters $\alpha,\overline{x}_i,\delta_i$.
Fig.~\ref{fig.sim2} illustrates the system response for different values of $w$. It can be observed in Fig.~\ref{fig.sim2}.(a) that the additive error induced by quantization does not vary significantly with $\xb(t)$ but increases as the communication bandwidth decreases, which is consistent with Theorem~\ref{thm:2}.

\subsection{Privacy Protection of Quantum Keys via Quantization}

\begin{figure}[t]
    \centering
    \begin{subfigure}[For different communication bandwidth $w$]{
\includegraphics[width=8.5cm]{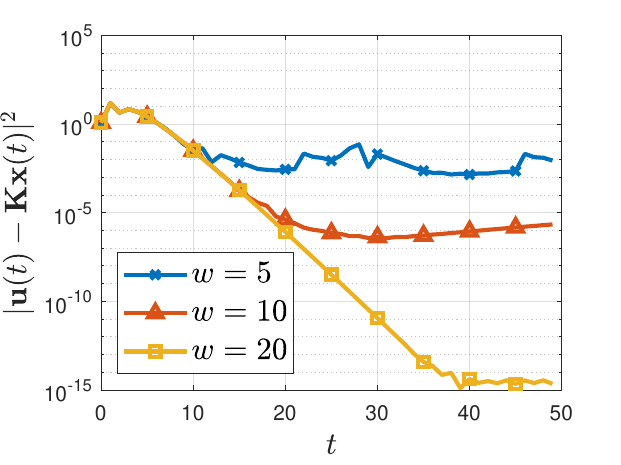}
       }
    \end{subfigure}
     \begin{subfigure}[For different privacy protection level $\Delta$]{
\includegraphics[width=8.5cm]{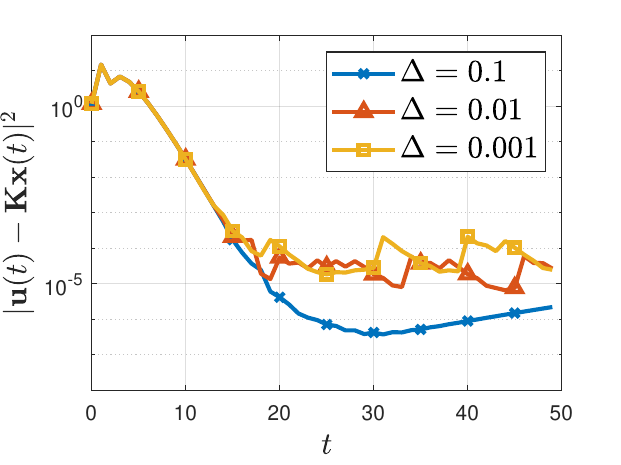}
       }
    \end{subfigure}
    \caption{Time evolutions   of $\|\ub(t)-\Kb\xb(t)\|^2$ in \eqref{eq:example_opendloop}-\eqref{eq:example_closedloop} encrypted by $\mathsf{QEC}$ \eqref{eq:beta_q}-\eqref{eq:dec2}.}
    \label{fig.sim2}
\end{figure}

We analyze how varying privacy levels affect the quantization error under a fixed communication bandwidth. To this end, we apply $\mathsf{QEC}$ defined by \eqref{eq:beta_q}-\eqref{eq:dec2} with a sufficiently large $w$, ensuring $(0,\Delta)$-differential privacy under a fixed $\zeta$-adjacency relation for different $\Delta$. As indicated by Theorem \ref{thm:4}, the privacy parameter $\Delta$ can be  modulated by adjusting $\alpha$. The results in Fig.~\ref{fig.sim2} show that stronger privacy protection (smaller $\Delta$) leads to a larger quantization  error, a trend consistent with the trade-off predicted in Corollary~\ref{cor:1}.

\subsection{Comparison with Classical Encryption Schemes}
We compare the proposed Exponential-Logarithmic Realization $\mathsf{QEC}$ \eqref{eq:enc}-\eqref{eq:dec} and several classical encrypted control schemes, including RSA~\cite{Con_HE1}, Paillier~\cite{PAR_CON}, and AES~\cite{A_AES}. In all cases, the communication bandwidth is kept identical, corresponding to $w = 10$ bits. The average time spent in each time step in the computation by the sensor, the controller, and the actuator is summarized in Table~\ref{tab:2},   {using MATLAB scripts running on an
 Intel Core i9-14900HX}. The computational efficiency of Exponential-Logarithmic Realization $\mathsf{QEC}$ \eqref{eq:enc}-\eqref{eq:dec} is clearly indicated.


\begin{table}[t]
\centering
\caption{Average computation time across encryption schemes ($10^{-5}$s)}
\begin{tabular}{l|cccc}
\toprule
{\bf Time in}& {\bf RSA} & {\bf Pailiar} & {\bf AES} & {\bf  QEC} \\
\midrule
{Sensor} & 272.0 & 281.6 & 4.2 & 0.8 \\
{Controller} & 1.4 & 6.2 & 10.0 & 0.6 \\
{Actuator} & 254.8 & 257.8 & 1.0 & 0.6 \\
{Total} & 528.2 & 545.6 & 15.2 & 2.0 \\
\bottomrule
\end{tabular}
\label{tab:2}
\end{table}

Next, we evaluate the confidentiality provided by these encryption schemes. In this scenario, we assume the eavesdropper adopts a common attack model: while it cannot obtain the key, it has access to the ciphertext of the system state \(\tilde{\xb}(t)\) for all \(t \in \mathbb{N}\), as well as the initial state \(\xb(0)\). Using the ciphertext, the eavesdropper applies the \texttt{n4sid} method in MATLAB for system identification, obtaining the matrix \(\Ab_{\rm e}\). Then, using the initial condition \(\xb(0)\), it simulates the system with the equation \(\xb_{\rm e}(t+1) = \Ab_{\rm e} \xb_{\rm e}(t)\) to obtain \(\xb_{\rm e}(t)\). In addition, we introduce various types of noise into the system, including Gaussian white noise, uniform noise, and impulse noise.
Table \ref{tab:3} shows the average relative error \(\frac{\|\xb_{\rm e}(t) - \xb(t)\|}{\|\xb(t)\|}\) across time steps, computed over 1000 repeated experiments. It is evident that the eavesdropper relying on system identification is unable to recover the original state of \(\xb(t)\),  demonstrating the confidentiality of the encrypted control system.

\begin{table}[t]
\centering
\caption{Average eavesdropping error across encryption schemes under different noises}
\begin{tabular}{l|ccccc}
\toprule 
{\bf Noises}& {\bf No Enc.} & {\bf RSA} & {\bf Pailiar} & {\bf AES} & {\bf  QEC} \\
\midrule
{No noise} & 0.00 & 0.82 & 1.52 & 1.53 & 1.60 \\
{Gaussian} & 0.08 & 1.05 & 1.53 & 1.54 &1.77 \\
{Uniform} & 0.20 & 1.10 & 1.18 & 0.79 & 0.91 \\
{Impulse} & 0.40 & 1.16 & 0.77 & 0.69 & 1.07 \\
\bottomrule
\end{tabular}
\label{tab:3}
\end{table}

\section{Conclusions}
\label{sec.con}
 
In this paper, we have introduced a quantum encrypted control framework for networked control systems, which was shown to deliver robust security with low computational overhead with the help of quantum channels. By eliminating the need for key distribution to the controller, \(\mathsf{QEC}\) reduces the risk of eavesdropping both in the communication channels and in the controller. Its security can be further ensured through quantum key distribution, making it resistant to attacks by quantum computers.
We analyzed the impact of quantum key errors on system stability and derived a critical threshold for quantum noise. Our results demonstrated that \(\mathsf{QEC}\) is resilient to key mismatches, ensuring stable system performance even under imperfect quantum key distribution. A stochastic quantizer is integrated into \(\mathsf{QEC}\) to address limited communication bandwidth and to guarantee privacy protection of the quantum key.

We believe this work opens up significant avenues for future research. First, alternative quantum encryption and decryption techniques could be explored, such as fully homomorphic encryption and advanced symmetric encryption, to further enhance both security and computational efficiency. Second, the proposed framework could be extended to more complex system dynamics, including output feedback control, LQR, MPC, and controllers for nonlinear systems, to evaluate its performance in more sophisticated control scenarios. Lastly, implementing \(\mathsf{QEC}\) in real-world cyber-physical systems, such as power grids or remotely controlled robots, would provide valuable insights into its scalability and effectiveness in real-time applications with large-scale networks and practical communication constraints.

\appendix
\section*{Appendix}

\section{Proof for  Theorem \ref{thm:3}}
\label{app:thm2}

For simplicity and without further repetition, in the following proof, the subscript \(i\) will be applied to \(i = 1, 2, \ldots, n\), and the index \(t\) to all \(t \in \mathbb{N}\).

Denote the parameter in \eqref{eq:beta} generated by $q_{\rm ac}(t)$ of the actuator as $\hat{\beta}_i(t)$. 
According to \eqref{eq:enc}–\eqref{eq:dec} and noting that $\beta_i(t) \neq 0$, we have  
\[
\ub(t) = \sum_{i=1}^n{\hat{\beta}_i(t)}\ln[\tilde{\ub}(t)]_i = \sum_{i=1}^n\frac{\hat{\beta}_i(t)}{\beta_i(t)}[\Kb]_i[\xb(t)]_i,\]
which yields
\begin{equation}
    \label{eq:eut}
    \ba
\wb(t)&=\sum_{i=1}^n\lambda_i(t)[\Kb]_i[\xb(t)]_i,
\ea
\end{equation}
for the intrinsic quantum error $\wb(t) := \ub(t) - \Kb \xb(t)$ and $\lambda_i(t) := \frac{\hat{\beta}_i(t)}{\beta_i(t)} - 1$.

Next, we analyze the expectation and variance of $\wb(t)$. 
By Lemma~\ref{propo}.(ii), we obtain  
\begin{equation}
    \label{eq:ed}
\begin{aligned}
\mathsf{E}(\hat{b}_{i,j}(t)|b_{i,j}(t)=B) =& B(1-p) + (1-B)p = B + (-1)^{B}p,\\
\mathsf{D}(\hat{b}_{i,j}(t)) =& \mathsf{E}(\hat{b}_{i,j}^2(t)) - \mathsf{E}^2(\hat{b}_{i,j}(t)) \\=&
b_{i,j}(t)^2(1-p)+(1-b_{i,j}(t))^2p-(b_{i,j}(t)+(-1)^{b_{i,j}(t)}p)^2\\=& p - p^2,
\end{aligned}
\end{equation}
for $B\in\{0,1\}$, $j=0,1,\dots,w_{b}-1$, where the fact $\hat{b}_{i,j}(t), b_{i,j}(t) \in \{0,1\}$ is used. 
Since $\hat{\beta}_i$ is a linear combination of the random variables $\hat{b}_{i,j}$, which are i.i.d. across all bit positions, it follows from \eqref{eq:dec} and \eqref{eq:ed} that  
\begin{equation}
\label{eq:ab}
\begin{aligned}
&\mathsf{E}(\hat{\beta}_i(t)|\beta_i(t)=\mathcal{B}) \\
=&\mathcal{B}+p\left(2b_{i,w_{b}-1}(t)(2^{w_b-1}+1)-\sum_{j=0}^{w_{b}-2}2b_{i,j}(t)2^j-2\right)\\
=&\mathcal{B}-2p\mathcal{B}
\end{aligned}
\end{equation}
for $\mathcal{B}\in\{-2^{w_b-1},-2^{w_b-1}+1,...,-1,1,..., 2^{w_b-1}\}$.
Then, we can derive the statistical properties of $\lambda_i(t)$ as  
\begin{equation}
 \label{eq:exp}
\ba
\mathsf{E}(\lambda_i(t)|\beta_i(t)=\mathcal{B})
&= \frac{1}{\mathcal{B}}\mathsf{E}(\hat{\beta}_i(t) - \mathcal{B}|\beta_i(t)=\mathcal{B})= 2p,
\ea
\end{equation}
which follows from \eqref{eq:ab} and \eqref{eq:enc}.
Furthermore, applying the law of total variance yields  
\begin{equation}
    \label{eq:var}
\begin{aligned}
\mathsf{D}(\lambda_i(t))&=\mathsf{E}(\mathsf{D}(\lambda_i(t)|\beta_i(t)))+\mathsf{D}(\mathsf{E}(\lambda_i(t)|\beta_i(t)))= \mathsf{E}\left(\frac{4^{w_b} + 3 \cdot 2^{w_b} + 2}{3}\frac{p-p^2}{\beta_i^2(t)}\right)+0=:h(p)
\end{aligned}
\end{equation}
where the second equality is obtained by substituting \eqref{eq:ab} and \eqref{eq:exp}.  
Since $\beta_i(t) \neq 0$, it follows that $0 \le h(p) < \infty$, and $h(p)$ is continuous with $h(0) = 0$.  
Moreover, as $\beta_i(t)$ is independent of both $i$ and $t$ by Lemma \ref{propo}.(i), the function $h(p)$ is also independent of $i$ and $t$.
Then, using \eqref{eq:exp}, \eqref{eq:var}, and \eqref{eq:eut}, we obtain
\begin{equation}
    \label{eq:fact2}
    \ba
\mathsf{E}(\wb(t)|\xb(t))&=2p\Kb\xb(t),\\
\mathsf{D}(\wb(t)|\xb(t))&=h(p)\|\Kb\xb(t)\|^2.
\ea
\end{equation}

Now we are ready to analyze the closed-loop system \eqref{eq:control}-\eqref{eq:controllaw}. We have
\[
\ba
\xb(t+1)\xb^\top(t+1)
=& \Mb_0\xb(t)\xb^\top(t)\Mb_0
+ \Mb_0\xb(t)\wb^\top(t)\Bb^\top
+ \Bb\wb(t)\xb^\top(t)\Mb_0
+ \Bb\wb(t)\wb^\top(t)\Bb^\top,
\ea
\]
where $\Mb_0 := \Ab + \Bb\Kb$.
Taking the conditional expectation with respect to $\xb(t)$ on both sides and applying \eqref{eq:fact2}, we obtain
\[
\ba
&\mathsf{E}(\xb(t+1)\xb^\top(t+1)|\xb(t))
\\=& \Mb_0\xb(t)\xb^\top(t)\Mb_0
+ \mathsf{E}(\wb(t)|\xb(t))\Mb_0\xb(t)\Bb^\top
+ \mathsf{E}(\wb(t)|\xb(t))\Bb\xb^\top(t)\Mb_0+ \mathsf{E}(\wb^2(t)|\xb(t)) \Bb\Bb^\top\\
=& \Mb_0\xb(t)\xb^\top(t)\Mb_0
+ 2p\Mb_0\xb(t)\xb^\top(t)\Mb_1^\top
+ 2p\Mb_1\xb(t)\xb^\top(t)\Mb_0 + \overline{h}(p)\Mb_1\xb(t)\xb^\top(t)\Mb_1^\top,
\ea
\]
where $\Mb_1 := \Bb\Kb$ and $\overline{h}(p) := 4p^2 + h(p)$.
Letting $\Vb(t) := {\rm vec}(\mathsf{E}(\xb(t)\xb^\top(t)))$ and taking the expectation of $\xb(t)$ of both sides, we can rewrite the above as
\[
\ba
\Vb(t+1)
=& \big(\Mb_0\otimes \Mb_0
+ 2p (\Mb_1\otimes \Mb_0 + \Mb_0\otimes \Mb_1)
+ \overline{h}(p)\Mb_1\otimes \Mb_1 \big)\Vb(t).
\ea
\]

Then, defining
\[
\ba
\Mb(p) :=& \Mb_0 \otimes \Mb_0 
    + 2p\,(\Mb_1 \otimes \Mb_0 + \Mb_0 \otimes \Mb_1)
    + \overline{h}(p)\,\Mb_1 \otimes \Mb_1,
    \ea
\]
it can be noticed that if we can show that $\rho(\Mb(p)) < 1$ for all $p \leq p^\ast$ with some $p^\ast > 0$, it follows that $\lim_{t\to\infty}\Vb(t)=0$, thereby completing the proof.
Since $\rho(\Mb_0) < 1$, there exists an induced matrix norm $\|\cdot\|_\ast$ such that $\|\Mb_0\|_\ast \leq  \rho(\Mb_0)+\frac{1}{2}(1-\rho(\Mb_0))=\frac{1}{2}(1+\rho(\Mb_0))<1$.  
Under this norm, we have
\begin{equation}
\label{eq:mp}
    \|\Mb(p)\|_\ast
    \leq \|\Mb_0\|_\ast^2
    + 4p\,\|\Mb_0\|_\ast\,\|\Mb_1\|_\ast
    + \overline{h}(p)\,\|\Mb_1\|_\ast^2.
\end{equation}
Define $\epsilon :=1-\rho(\Mb_0)$. For
\[
\ba
p \leq p^\ast =& \frac{\epsilon}{\max\{16\|\Mb_1\|^2_\ast,16\|\Mb_1\|_\ast,4A\|\Mb_1\|^2_\ast\}},
\ea
\]
where $A:=\frac{4^{w_b} + 3 \cdot 2^{w_b} + 2}{3\mathsf{E}(\beta_i^2(t))}$,
we have
\[
\ba
\overline{h}(p)&=h(p)+4p^2\leq A(p-p^2)+4p^2\leq \frac{\epsilon}{4\|\Mb_1\|^2},
\ea
\]
where the fact $p^2\leq p$ is used. Then by \eqref{eq:mp}, it can be
obtained
\begin{equation}
\|\Mb(p)\|_\ast
<  \|\Mb_0\|_\ast^2 + \frac{\epsilon}{4}+ \frac{\epsilon}{4}
 \leq 1,
\end{equation}
where the first inequality uses the $\|\Mb_0\|_\ast < 1$.
Finally, since $\rho(\Mb(p)) \le \|\Mb(p)\|_\ast$, the result $\lim_{t\to\infty}\Vb(t)=0$ follows, completing the proof.

\section{Proof for  Theorem \ref{thm:2}}
\label{app:thm3}

In the following, for simplicity, we discuss the validity of the conclusion at an arbitrary specific time \( t\in\mathbb{N} \) and omit the time index \((t)\) of variables. In addition, without further repetition, in the following proof, the subscript \( i \) applies to all \( i = 1, 2, \ldots, n \).

From \eqref{eq:dec2}, we have
\[
\ub 
= \sum_{i=1}^n\beta_i \ln(z_i)
   + \sum_{i=1}^n\delta_i \beta_i
     \ln\!\left(\frac{g^{-1}\left(\sum_{j=0}^{w-1} 2^{-j} d_{i,j}\right)}{z^{1/\delta}}\right),
\]
which yields
\begin{equation}
\label{eq:q_u}
\ba
&\mathsf{E}(\ub - \Kb \xb)^2
\\\le& 2\mathsf{E}\left( \sum_{i=1}^n\beta_i\ln(z_i) - \sum_{i=1}^n [\Kb]_i [\xb]_i \right)^2
+ 2\mathsf{E}\left(\sum_{i=1}^n\delta_i \beta_i
     \ln\!\left(\frac{g^{-1}\left(\sum_{j=0}^{w-1} 2^{-j} d_{i,j}\right)}{z^{1/\delta}}\right)\right)^2.
\ea
\end{equation}

In the following, we first establish the upper bound of the first term in \eqref{eq:q_u}. With \eqref{eq:beta_q}, it can be obtained that $|\beta_i|\geq \frac{|[\xb]_i|}{\ln(1+\alpha)}$ by \eqref{eq:beta_q}, $\overline{x}_i>{|[\xb(t)]_i|}$ and $\alpha\in(0,1]$. Then we have
\begin{equation}
    \label{eq:vibound}
\ba
& \frac{1}{2} \le v_i = \exp({[\xb]_i/\beta_i}) \le 2. 
\ea
\end{equation}
It can be noted that, for $v\in\left[\frac{1}{2},2\right]$ and $\overline{a_{w-1}a_{w-2}...a_0}=Q_w(v)$, there holds 
\begin{equation}
    \label{eq:E2}
    \ba
\mathsf{E}\left(g^{-1}\left(\sum_{j=0}^{w-1}2^{-j}a_{j}\right)-v\right)^2&\leq\mathsf{E}\left(\sum_{j=0}^{w-1}2^{-j}a_{j}-g(v)\right)^2\leq \frac{1}{2^{2w}},
\ea
\end{equation}
where the first inequality is obtained by noting $g^{-1}(y)$ is $1$-Lipschitz for $y\in[0,2]$, and the last inequality is obtained by \eqref{eq:qw}.
 From \eqref{eq:qw0}, \eqref{eq:vibound} and \eqref{eq:E2}, we obtain
\begin{equation}
\label{eq:q_ei_0}
\mathsf{E}(e_i)  =0,
\end{equation}
\begin{equation}
\label{eq:q_ei}
\mathsf{E}|e_i |^2 \le \frac{1}{2^{2w}},
\end{equation}
for $
e_i := g^{-1}(\sum_{j=0}^{w-1} 2^{-j} c_{i,j}) - v_i.
$
By \eqref{eq:con2}, we have
\[
\ba
z_i &=  \left(v_i + e_i\right)^{ [\Kb]_i} =  v_i^{[\Kb]_i} 
      \left(1 + \frac{e_i}{v_i}\right)^{[\Kb]_i} = z^r_i  \left(1 + \frac{e_i}{v_i}\right)^{[\Kb]_i}
\ea
\]
for $z^r_i := \exp{({ [\Kb]_i [\xb]_i / \beta}_i)}$, where the last equality is obtained by  \eqref{eq:enc2}.
By Taylor expansion, we have
\begin{equation}
\label{eq:q_z2}
\frac{z_i}{z^r_i} - 1 
= [\Kb]_i\frac{e_i}{v_i}+\frac{[\Kb]_i([\Kb]_i-1)}{2}\left(\frac{e_i}{v_i}\right)^2+o\left(\frac{e_i^2}{v_i^2}\right).
\end{equation} 
Then, for the first term in \eqref{eq:q_u}, we have
\[
\ba
&\mathsf{E}\left( \sum_{i=1}^n\beta_i\ln(z_i)  - \sum_{i=1}^n [\Kb]_i [\xb]_i \right)^2
\\=& \mathsf{E}\left(\sum_{i=1}^n\left( \beta_i \ln(z_i) - \beta_i \ln(z^r_i) \right)\right)^2
\\=& \mathsf{E}\left( \sum_{i=1}^n\beta_i \ln\left(1+\frac{z_i}{z_i^r}-1\right)  \right)^2
\\=&  \sum_{i=1}^n\mathsf{E}\left(\beta_i \ln\left(1+\frac{z_i}{z_i^r}-1\right)  \right)^2+2\sum_{i<j}\mathsf{E}\left(\beta_i \ln\left(1+\frac{z_i}{z_i^r}-1\right)\right)\mathsf{E}\left(\beta_j \ln\left(1+\frac{z_j}{z_j^r}-1\right)\right)\\
=& \sum_{i=1}^n \mathsf{E}\left(\left(\beta_i
   [\Kb]_i\frac{e_i}{v_i}\right)^2
   +o\left(\frac{e_i^2}{v_i^2}\right)
\right),
\ea
\]
where the last  equality follows from  \eqref{eq:q_z2}, Taylor expansion of $\ln(1+x)$ at $x=0$ and \eqref{eq:q_ei_0}. Then we have
\begin{equation}
\label{eq:ft}
\ba
&\mathsf{E}\left( \sum_{i=1}^n\beta_i\ln(z_i)  - \sum_{i=1}^n [\Kb]_i [\xb]_i \right)^2
\leq \frac{\sum_{i=1}^n (\beta_i\delta_i)^2}{2^{2w-2}} 
  + o\!\left(\frac{1}{2^{2w}}\right),
\ea
\end{equation}
where we use $v_i\geq\frac{1}{2}$, \eqref{eq:q_ei} and $\delta_i\geq |[\Kb]_i|$.

Next, we establish the upper bound of the second term in \eqref{eq:q_u}. By \eqref{eq:con2}, it follows that
\[
2^{-|[\Kb]_i|} \le z \le 2^{ |[\Kb]_i|}.
\]
Since $\delta _i\geq |[\Kb]_i|$, we obtain $\frac{1}{2} \le z_i^{1/\delta_i} \le 2$.
Hence, by \eqref{eq:qw0}, \eqref{eq:E2} and \eqref{eq:con2}, we have
\begin{equation}
\label{eq:q_fact2}
\ba
&\mathsf{E}\left(g^{-1}\left(\sum_{j=0}^{w-1} 2^{-j} d_{i,j} \right)- z^{1/\delta_i}\right)
=0,\\
&\mathsf{E}\left(g^{-1}\left(\sum_{j=0}^{w-1} 2^{-j} d_{i,j} \right)- z^{1/\delta_i}\right)^2
\le \frac{1}{2^{2w}}.
\ea
\end{equation}
Then, similar to \eqref{eq:ft}, we directly obtain
\begin{equation}
\label{eq:st}
\ba
&\mathsf{E}\left(\sum_{i=1}^n\delta_i \beta_i 
\ln\!\left(1+\frac{\sum_{j=0}^{w-1} 2^{-j} d_{i,j} }{z^{1/\delta_i}}-1\right)\right)^2
\leq \sum_{i=1}^n\frac{(\delta_i \beta_i )^2}{2^{2w}}+o\!\left(\frac{1}{2^{2w}}\right),
\ea
\end{equation}
where the inequality is obtained by Taylor expansion and \eqref{eq:q_fact2}.

Finally, substituting \eqref{eq:ft} and \eqref{eq:st} into \eqref{eq:q_u} yields the desired result.

\section{Proof for Theorem \ref{thm:4}}
\label{app:thm4}

First, we propose the following auxiliary Lemma for $h(\cdot)$ defined in \eqref{eq:hde}.
\begin{lemma}
\label{lem:aux}
    Let $(\mathbb{R},\mathcal{F}_0 , \mathbb{P})$ be a probability space. Then, for all $y,y'\in[0,2]$ and all $s\in[0,\frac{1}{2^{w-1}}]$ satisfying $|y-y'|\leq s$, there holds
    \[
    |\mathbb{P}(h_w(y)\in\mathbb{S})- \mathbb{P}(h_w(y')\in\mathbb{S})|\leq {2^{w-1}}s
    \]
    for all $\mathbb{S}\in\mathcal{F}_0$.
\end{lemma}
The proof of Lemma \ref{lem:aux} can be obtained directly  according to Lemma 1.1 in \cite{Liu2025Design}.

According to the function $g(v)$ defined in \eqref{eq:vfun}, we know for all $v,v'\in\left[\frac{1}{2},2\right]$ satisfying $|v-v'|\leq s$, there holds $y,y'\in[0,2]$ and $|y-y'|\leq s$ for $y=g(v)$ and $y'=g(v')$. From Lemma \ref{lem:aux}, we know for such $v,v'$, if $s\in[0,\frac{1}{2^{w-1}}]$ there holds
 \begin{equation}
     \label{eq:con1}
    |\mathbb{P}(Q_w(v)\in\mathbb{S})- \mathbb{P}(Q_w(v')\in\mathbb{S})|\leq {2^{w-1}}s
 \end{equation}
    for all $\mathbb{S}\in\mathcal{F}_0$, where we use the fact $Q_w(x)=h_w(g(x))$.

For any fixed $i\in\{1,2,...,n\}$ and $t\in\mathbb{N}$, we assume $\beta_i(t),\beta_i'(t)$  defined in \eqref{eq:beta_q} satisfy \eqref{eq:beta_d} for $\zeta>0$. Then, for $v_i(t)=\exp\left(\frac{[\xb(t)]_i}{\beta_i(t)}\right)$ and $v_i'(t)=\exp\left(\frac{[\xb(t)]_i}{\beta_i'(t)}\right)$, there holds 
\begin{equation}
    \label{eq:vv'}
\ba
&|v_i(t)-v_i'(t)|\leq  \zeta \sup_{|\beta|\geq\frac{\overline{x}_i}{\ln(1+\alpha)} }\frac{{\rm d}(\exp\left(\frac{[\xb(t)]_i}{\beta}\right))}{{\rm d}\beta}\leq \zeta\frac{(1+\alpha)\ln^2(1+\alpha)}{\overline{x}_i},
\ea
\end{equation}
where the first inequality uses the fact $|\beta_i(t)|\geq \frac{\overline{x}_i}{\ln(1+\alpha)}$ by \eqref{eq:beta_q} and the last inequality is obtained by $\overline{x}_i\geq |[\xb(t)]_i|$. Next, assume  $\zeta\frac{(1+\alpha)\ln^2(1+\alpha)}{\overline{x}_i}\in[0,\frac
{1}{2^{w-1}}]$. From \eqref{eq:con1} and \eqref{eq:vv'}, we know for $\beta_i(t),\beta_i'(t)$   satisfying \eqref{eq:beta_d} with $\zeta>0$, $v_i(t)=\exp\left(\frac{[\xb(t)]_i}{\beta_i(t)}\right)$ and $v_i'(t)=\exp\left(\frac{[\xb(t)]_i}{\beta_i'(t)}\right)$ satisfy that
\begin{equation}
     \label{eq:conc2}
     \ba
    &|\mathbb{P}(Q_w(v_i(t))\in\mathbb{S})- \mathbb{P}(Q_w(v_i'(t))\in\mathbb{S})|\leq\zeta\frac{(1+\alpha)\ln^2(1+\alpha)}{\overline{x}_i}{2^{w-1}},\quad \forall\ \mathbb{S}\in \mathcal{F}_0.
    \ea
 \end{equation}

In $O_T(\mathbf{E}_T)$, the element $[\tilde{\xb}(t^\ast)]_{i^\ast}$ is the only one that directly depends on $\beta_{i^\ast}(t^\ast)$. All other elements are independent of $\beta_{i^\ast}(t^\ast)$ or depend on $\beta_{i^\ast}(t^\ast)$ only through $[\tilde{\xb}(t^\ast)]_{i^\ast}$. Hence,  by the post-processing property of differential privacy, the privacy guarantee for the entire trajectory is determined by that of $[\tilde{\xb}(t^\ast)]_{i^\ast}$, i.e.,
\[
\ba
&\sup_{\mathcal{\mathbb{O}\in\mathcal{F}}}|\mathbb{P}(O_T(\mathbf{E}_T)\in\mathbb{O})-\mathbb{P}(O_T(\mathbf{E}_T'))\in\mathbb{O})|
\leq \sup_{\mathcal{\mathbb{S}\in\mathcal{F}}_0}|\mathbb{P}([\tilde\xb_{i^\ast}(t^\ast)]\in\mathbb{S})- \mathbb{P}([\tilde\xb_{i^\ast}'(t^\ast)]\in\mathbb{S})|
\leq  \zeta\frac{(1+\alpha)\ln^2(1+\alpha)}{\overline{x}_{i^\ast}}{2^{w-1}},
\ea
\]
where the last inequality is obtained by \eqref{eq:conc2} and \eqref{eq:enc2}. Then we complete the proof.

\section{Proof for Corollary \ref{cor:1}}
\label{app:cor1}
It can be noted that the mean square of $\ub(t)-\Kb\xb(t)$ in Theorem \ref{thm:2} can be further bounded by 
\[
    \mathsf{E}\left( \ub(t) - \Kb \xb(t) \right)^2
    \leq 
    \frac{5n(2^{w_b-1}+\frac{\overline{x}}{\ln(1+\alpha)})^2 \delta^2}{2^{2w}},
\]
where   ${\delta}:=\max_{i=1,2,...,n}{\delta}_i$, and the higher-order small quantity is omitted.
Then, defining $\alpha_1= e^{\overline{x}/2^{w_b-1}}-1$, we know if $\alpha\in(0,\alpha_1]$, there holds 
\[
    \mathsf{E}\left( \ub(t) - \Kb \xb(t) \right)^2
    \leq 
    \frac{20n{\overline{x}^2} \delta^2}{2^{2w}\ln^2(1+\alpha)},
\]
which yields $\mathsf{E}\left( \ub(t) - \Kb \xb(t) \right)^2\leq E^2_g$ if $\alpha\geq \alpha_2=\frac{2\sqrt{5n}\overline{x}{\delta}}{2^w E_g}$.

On the other hand, from Theorem \ref{thm:4}, it can be obtained that $\mathsf{QEC}$ is $(0, \Delta_g)$-differentially private under the $\zeta_g$-adjacency relation if  
  \[
   \alpha\leq\alpha_3= e^{\sqrt{{\Delta_g\overline{x}}/{(2^w\zeta_g)}}}-1.
  \]
Then, if 
\[
w \ge \max\left\{
w_b + 1 + \log_2\left(\frac{\sqrt{5n}\,{\delta}}{E_g}\right),\;
\left(\log_2\frac{20n \overline{x} {\delta}^2 \Delta_g}{E_g^2\zeta_g}
\right)\right\},
\]
we have $0<\alpha_2<\min\{\alpha_1,\alpha_3\}$. In this case, the corollary  holds with  $\alpha\in[\alpha_2,\min\{\alpha_1,\alpha_3\}]$. We have completed the proof. 

\section*{Acknowledgment}
The authors thank Elizabeth Ratnam (Monash University) and Farhad Farokhi  (The University of Melbourne) for insightful discussions that inspired the early ideas of this paper.

\end{document}